\documentclass[twocolumn,twocolappendix]{aastex631}

\usepackage{amsmath}
\usepackage{cjhebrew}
\usepackage{lipsum}
\usepackage{float}
\usepackage{afterpage}
\usepackage{makecell}
\usepackage{upgreek}
\usepackage{bm}
\usepackage{ulem}
\usepackage{starfont}
\usepackage[T1]{fontenc}
\usepackage{calligra}

\graphicspath{{./}{figures/}}

\newcommand{\tensorSymbol}[1]{{\overset{\leftrightarrow}{#1}}}

\graphicspath{{./}{figures/}}
\shorttitle{Accreting Unequal-Mass Binaries}
\shortauthors{Dittmann \& Ryan}

\begin{document}

\title{The Evolution of Accreting Binaries: from Brown Dwarfs to Supermassive Black Holes}

\correspondingauthor{Alexander J. Dittmann}
\email{dittmann@astro.umd.edu}

\author[0000-0001-6157-6722]{Alexander J.~Dittmann}
\affil{Department of Astronomy and Joint Space-Science Institute, University of Maryland, College Park, MD 20742-2421, USA}
\affil{Theoretical Division, Los Alamos National Laboratory, Los Alamos, NM 87545, USA}

\author[0000-0001-9068-7157]{Geoffrey Ryan}
\affil{Perimeter Institute for Theoretical Physics, 31 Caroline St. N., Waterloo, ON, N2L 2Y5, Canada}

\begin{abstract} 
Circumbinary accretion occurs throughout the universe, from the formation of stars and planets to the aftermath of major galactic mergers.
We present an extensive investigation of circumbinary accretion disks, studying circular binaries with mass ratios ($q\equiv M_2/M_1$) from 0.01 to 1 and at each mass ratio probing the effects of disk thickness and viscosity. We study disks with aspect ratios $H/r\in\{0.1, 0.05, 0.033\}$, and vary both the magnitude and spatial dependance of viscosity. Although thin accretion disks have previously been found to promote rapid inspirals of equal-mass binaries, we find that gravitational torques become weaker at lower mass ratios and most binaries with $0.01\leq q\leq0.04$ outspiral, which may delay the coalescence of black hole binaries formed from minor mergers and cause high-mass exoplanets to migrate outwards. However, in a number of cases, the disks accreting onto binaries with mass ratios $\sim 0.07$ fail to develop eccentric modes, leading to extremely rapid inspirals. Variability in black hole accretion correlates with disk eccentricity, and we observe variability above the $\sim10\%$ level even for mass ratios of $0.01$. We demonstrate that the spatial dependence of the viscosity (e.g. $\alpha$ vs constant-$\nu$) significantly affects the degree of preferential accretion onto the secondary, resolving discrepancies between previous studies. 
Colder circumbinary disks remain eccentric even at $q\sim0.01$ and sustain deep, asymmetric cavities. 
\end{abstract}

\section{Introduction} Circumbinary accretion disks, in a general sense, occur whenever a binary of sufficient mass interacts with a surrounding gaseous disk. When the binary members are of comparable mass, such systems arise, for example, during star formation \citep[e.g.,][]{1986ApJS...62..519B,1997MNRAS.285...33B,2022arXiv221204564P}, where interactions with between the disk and binary may shape the architecture of observed stellar systems \citep{2019MNRAS.489.5822E,2022ApJ...933L..32H}; and after the merger of comparable-mass galaxies, where interactions with gaseous disks may expedite supermassive black hole coalescence \citep[e.g.,][]{1980Natur.287..307B,2020ApJ...900...43T,2022MNRAS.513.6158D}. Circumbinary accretion systems containing binaries of more disparate masses can result from, for example, the formation of high-mass planets in protoplanetary disks \citep[e.g.,][]{2004A&A...425L..29C,2018NatAs...2..138B}, and the accretion of gas onto the binaries resulting from the merger of massive and dwarf galaxies.

Binary black holes are likely to form in the centers of galaxies following mergers, and gravitational interactions between a binary and the surrounding gas and stars may help these black holes reach sufficiently small separation that the binary can be driven to merge by gravitational radiation \citep[e.g.,][]{1980Natur.287..307B,2009ApJ...700.1952H,2013ApJ...773..100K}. While major mergers typically result in the formation of binaries with comparable masses, the sheer abundance of dwarf galaxies across cosmic time suggests that mergers between galaxies of greatly disparate masses should be much more common \citep[e.g.,][]{2000MNRAS.318L..35H,2007MNRAS.374....2R}. Although the orbital evolution of binaries formed from mergers involving dwarf galaxies may be complicated by factors such as the dark matter distributions in the merging galaxies \citep{2018ApJ...864L..19T}, tidal stripping \citep{2003MNRAS.341..434T}, and three-body dynamics \citep{2010MNRAS.402.2308A}, the binaries formed from such galaxy mergers may dominate the population of black hole mergers detected by LISA \citep[e.g.,][which suggest a peak at $q\sim0.05$]{2019MNRAS.482.2913B,2022arXiv220306016A}. Even if gas does not play a dynamically significant role in the orbital evolution of these binary systems, binary-driven variability may be detectable in these systems, as well as presently studied binary AGN candidates (e.g. \citet{2015MNRAS.453.1562G,2016MNRAS.463.2145C,2020MNRAS.499.2245C}, but see \citet{2016ApJ...833....6L,2016MNRAS.461.3145V,2018ApJ...859L..12L} for cautionary remarks). Additionally, appreciable alterations of binary black hole masses and orbital parameters can appreicably modify the spectrum of the stochastic gravitational wave background \citep[e.g.,][]{1995ApJ...446..543R,2001astro.ph..8028P,2020MNRAS.498..537S,2021MNRAS.507.1458F,2023arXiv230616213A}.

On more local scales, as the population of detected exoplanets with high mass ratios grows
we can begin to glean insights into different populations thereof. For example, although gravitational instabilities might produce large-mass large-separation directly imaged exoplanets \citep[e.g.,][]{2018haex.bookE.155B}, their mass distribution may also be consistent with the core accretion scenario \citep{2019AJ....158...13N,2019ApJ...877...46W}. Additionally, the factors that limit accretion onto high-mass planets are also uncertain but of great interest \citep[e.g.,][]{2020MNRAS.498.2054R,2021ApJ...906...52L}, as these affect the upper limits on planetary masses. 

The problem of accretion disks around unequal-mass binaries has been investigated numerous times \citep[e.g.,][]{1996ApJ...467L..77A,1997MNRAS.285...33B,2007PASJ...59..427H,2015ApJ...807..131S,2016MNRAS.459.2379D,2020ApJ...889..114M,2020MNRAS.499.3362R,2022A&A...660A.101P,2023arXiv230501533M,2023arXiv230801967C}. Although recent studies of thin disks have found nearly universally that the lower-mass binary member accretes preferentially \citep[e.g.,][]{2014ApJ...783..134F,2020ApJ...901...25D, 2021ApJ...921...71D,2022arXiv220302514S}, the precise degree of preferential accretion (measured as the ratio of the accretion rate onto the secondary to that onto the primary) at a given mass ratio has been found to vary by factors of $\sim3$ even between studies utilizing different versions of the same code \citep[e.g.,][]{2020ApJ...901...25D, 2021ApJ...921...71D}, due at least in part to different treatments of accretion and viscosity.
Additionally, studies of accretion onto binaries from thick disks have observed preferential accretion instead onto the primary, behavior which has been attributed to the effects of gas pressure on the dynamics of gas within the Hill sphere of the secondary \citep[c.f.][]{2005ApJ...623..922O,2010ApJ...708..485H,2015MNRAS.447.2907Y}.

Herein we present a systematic survey of accretion onto binaries of unequal masses. We study binaries with mass ratios $q\in[0.01,0.02,0.03,0.04,0.05,0.07,0.1,0.14,0.2,0.3,0.5,1]$. For each mass ratio, we simulate disks with viscosities $\nu=0.001\Omega_ba_b^2$ and $\mathcal{M}=\{10,20,30\}$, where $\Omega_b$ is the orbital angular frequency of the binary, $a_b$ its semi-major axis, and $\mathcal{M}$ is the characteristic azimuthal Mach number of the disk. At $\mathcal{M}=10$ we additionally simulate disks with viscosities of $\nu=0.0005\Omega_ba_b^2$ and $\alpha-$viscosities, setting $\alpha=0.1$. We describe our numerical methodology and simulation diagnostics in Section \ref{sec:methods}. We present the main results of our study in Section \ref{sec:results}, discuss their relevance to various astrophysical scenarios in Section \ref{sec:discussion}, and summarize our results in Section \ref{sec:summary}. We study the convergence of our simulations in Appendix \ref{app:convergence}, demonstrating not only that our results are generally converged with respect to both numerical resolution and mass removal rate, but also the pernicious effects of overly simple sink prescriptions. Appendix \ref{app:compare} makes direct comparisons between our work and two prior investigations, \citet{2020ApJ...901...25D} and \citet{2021ApJ...918L..36D}, in the low-mass ratio regime. Appendix \ref{app:summary} collects the quantitative results of our main suite of simulations. 

\section{Methods}\label{sec:methods}
This study primarily utilizes \texttt{Athena++} \citep{2020ApJS..249....4S},\footnote{\url{https://github.com/PrincetonUniversity/athena}} an Eulerian code, to solve the compressible Navier-Stokes equations, given by
\begin{align}
\partial_t\Sigma + \bm{\nabla}\!\cdot\!(\Sigma\mathbf{v}) &= S_\Sigma + D_\Sigma \label{eq:continuity} \\
\partial_t(\Sigma\mathbf{v}) + \bm{\nabla}\!\cdot\!(\Sigma\mathbf{v}\mathbf{v}+\Pi\tensorSymbol{I} - 2 \Sigma \nu \tensorSymbol{\sigma})
&= -\Sigma\bm{\nabla}\Phi + \mathbf{S}_{p} + \mathbf{D}_{p} \label{eq:momentum} 
\end{align}
$\Sigma$ is the disk surface density, $\Pi$ is the vertically-integrated pressure, $\mathbf{v}$ is the fluid velocity vector, $\Phi$ is the gravitational potential, $\tensorSymbol{I}$ is the identity matrix, $S_\Sigma$ is a mass sink term, $\mathbf{S}_{p} $ is a momentum sink term, $\tensorSymbol{\sigma}$ is the velocity shear tensor, $\nu$ is the kinematic viscosity, and additional damping source terms are given by $D_\Sigma$ and $\mathbf{D}_{p}$, the details of which are presented in Section \ref{sec:setup}. 

We implement a `locally isothermal' equation of state similarly to \citet{2019ApJ...875...66M}
by setting the adiabatic index of the fluid to $\Gamma=1.0001$ and implementing a source term which instantaneously sets $\Pi$ in each cell consitently with a prescribed local sound speed $c_s$:
\begin{equation}
\Pi=c_s^2\Sigma/\Gamma.
\end{equation}
The squared sound speed is given by 
\begin{equation}
c_s^2(\mathbf{x}) = -\frac{\Phi(\mathbf{x})}{\mathcal{M}^2},
\end{equation}
where $\mathcal{M}$ is a fiducial Mach number which we hold fixed in each simulation. 
We use a softened gravitational potential, where the total potential is given by $\Phi=\sum_i \Phi_i$
and $\Phi_i$ is the gravitational potential of a point mass of mass $M_i$ at position $\mathbf{x}_i$ given by
\begin{equation}\label{eq:plummer}
\Phi_i = \frac{-GM_i}{\sqrt{|\textbf{x}-\textbf{x}_i|^2+\epsilon_{g,i}^2}},
\end{equation}
where $\epsilon_{g,i}$ is a gravitational softening length. We hold the binary orbit fixed, at a semi-major axis $a_b$ and mass $M=M_1+M_2$, over the course of each simulation, calculating positions analytically each time step, which is justified as long as evolutionary timescales are many times longer than the binary orbital period and simulation duration. 

We consider multiple kinematic viscosity ($\nu$) prescriptions, including a globally constant kinematic viscosity and an $\alpha$-viscosity ansatz \citep{1973A&A....24..337S} such that $\nu=\alpha c_s H$, where $H$ is the disk scale height which we calculate using
\begin{equation}\label{eq:alphavisc}
H = \frac{c_s}{\sqrt{\sum_i GM_i(|\textbf{x}-\textbf{x}_i|^2+\epsilon_g^2)^{-3/2}}},
\end{equation}
which is appropriate to a thin disk in vertical hydrostatic equilibrium about multiple point masses with the gravitational potentials given by Equation (\ref{eq:plummer}). Because the degree of preferential accretion onto the secondary depends on the functional form of the viscosity, we test an additional viscosity prescription with an even stronger distance dependence than that of an $\alpha$-viscosity in Section \ref{sec:viscpref}.

The surface density sink term is given by 
\begin{equation}\label{surfSink}
S_\Sigma = -\gamma_i \Omega_b \Sigma \sum_i s_i(|\textbf{x}-\textbf{x}_i|),
\end{equation}
where $s_i$ is a function specifying the sink profile for each particle, $\Omega_b$ is the angular frequency of the binary, and $\gamma_i$ is the sink rate for each particle. Throughout this work, we use the sink profile
\begin{equation}\label{eq:GaussSquared}
s_i = \exp\left(-\frac{|\mathbf{x}-\textbf{x}_i|^4}{r_{s,i}^4}  \right),
\end{equation}
where the sink radius for each particle $r_{s,i}$ is a purely numerical parameter defining a characteristic radius over which the sink term acts.

We use a torque-controlled momentum sink term, given by
\begin{align} 
    \mathbf{S}_{p} &= -\gamma_i \Omega_b \Sigma \sum_i s_i(|\textbf{x}-\textbf{x}_i|)\mathbf{v}^*_i \label{eq:tf1} \\
    \mathbf{v}^*_i &= \left(\mathbf{v} - \mathbf{v}_i\right)\cdot \left(\hat{\mathbf{r}}_i \hat{\mathbf{r}}_i + \delta\ \! \hat{\bm{\phi}}_i \hat{\bm{\phi}}_i \right) + \mathbf{v}_i, \label{eq:tf2}
\end{align}
where $\mathbf{v}_i$ is the velocity of the sink particle,  $\hat{\mathbf{r}}_i$ and $\hat{\bm{\phi}}_i$ are the unit basis vectors at position $\mathbf{x}$ of a polar coordinate system centred on the sink particle, and $\delta$ is a dimensionless control parameter. Throughout most of this work, we set $\delta=0$, which results in a sink term that is torque-free in the frame of a given sink particle, leading to no change in the spin angular momentum of this sinks. Such sink methods make simulation results more robust to numerical parameters such as $\gamma$ \citep{2021ApJ...921...71D}, and are able to reproduce analytic steady-state disk profiles in single-object disks \citep{2020ApJ...892L..29D,2021ApJ...921...71D}. However, in Appendix \ref{app:convergence}, we present the results of a few simulations which used $\delta=1$, which sets $\mathbf{v}^*=\mathbf{v}$, to illustrate how this choice intrinsically leads to spurious results that depend strongly on the sink rate ($\gamma$).

\subsection{Diagnostics}\label{sec:diag}
Over the course of each simulation, we record the matter and angular momentum accreted by each binary component, as well as the angular momentum exchanged between each member of the binary and the surrounding gas through gravitational interactions. We integrate these quantities on every time step over the entire simulation, and output these volume- and time-integrals at a cadence of 100 outputs each binary orbital period.  

The accretion rate onto each particle is given by 
\begin{equation}
    \dot{M}_i = - \int\! dA\ S_{\Sigma, i}\ ,
\end{equation}
where $S_{\Sigma, i} = -\gamma \Omega_b \Sigma s_i(|\mathbf{x}-\mathbf{x}_i|)$. We define the momentum sink term for each particle analogously as $\mathbf{S}_{p, i} = S_{\Sigma_i} \mathbf{v}_i^*$ via Equation \eqref{eq:tf1}. Although these integrals formally extend over the entire domain, the sink profile function $s_i(x)$ sharply truncates contributions from beyond more than a few $r_s$ from a given sink particle. 
The torque on each particle due to accretion $\dot{J}_a$ is then given by
\begin{align}
    \dot{J}_{a,i} &= - \int \! dA\ \mathbf{x} \times \mathbf{S}_{p, i} = - \int \! dA\ S_{\Sigma, i}\ \mathbf{x} \times \mathbf{v}^*_{i}\ .
\end{align}
The change in angular momentum of each particle is computed every time step using
\begin{equation}
    \dot{J}_{g,i} =  -\int \! dA\ \mathbf{x} \times \left(-\Sigma \bm{\nabla} \Phi_i\right) 
\end{equation}

The total torque on the system of sink particles is the sum of the torques on each component:
\begin{equation}
    \dot{J} = \sum_i \dot{J}_{g, i} + \dot{J}_{a,i},
\end{equation}
which is also the rate of change of the binary orbital angular momentum because we employ (spin) torque-free sink particles.

We calculate the orbital evolution of the binary from the torques measured in our simulations using the equation for the binary orbital angular momentum
\begin{equation}
J_b=\frac{M_1M_2}{M}\sqrt{GMa_b},
\end{equation}
where $M$ is the total mass of the binary and we have specialized to circular binaries. Previous studies have found that small eccentricities $e_b \lesssim 0.08$ are damped through interaction with circumbinary disks \citep{2019ApJ...871...84M,2020arXiv201009707Z}, so this approximation is appropriate for binaries which form with low initial eccentricities. 

The torque on the binary can be expressed as 
\begin{equation} \label{eq:evolEqn1}
\frac{\dot{J}_{\rm orb}}{J_b} = \frac{\dot{M}_1}{M_1} + \frac{\dot{M}_2}{M_2} - \frac{1}{2}\frac{\dot{M}}{M} + \frac{1}{2}\frac{\dot{a}_b}{a_b},
\end{equation}
from which an expression for the evolution of the binary semi-major axis follows:
\begin{align}
\frac{d\log{a_b}}{d\log{M}} = 1+ 2\left[\frac{l_0}{
l_b}-\frac{\dot{M}_1}{\dot{M}}(1+q)-\frac{\dot{M}_2}{\dot{M}}\frac{1+q}{q}\right]\\=2\left[\frac{l_0}{l_b}-\left(\frac{1+q}{1+\lambda}\right)\left(1+\frac{\lambda}{q}\right)+\frac{1}{2}\right], \label{eq:evolEqn2}
\end{align}
where $q=M_2/M_1$ is the mass ratio, $\lambda\equiv\dot{M}_2/\dot{M}_1$ is the degree of preferential accretion onto the secondary, $l_0=\dot{J}_{\rm orb}/\dot{M}$ is the $\textit{orbital}$ angular momentum change in the binary per unit mass, and $l_b$ is the specific angular momentum of the binary $\Omega_ba^2_bq/(1+q)^2$. If gas is able to circularize before accreting, then the contribution to the overall torque due to accretion is given by $\dot{J}_{a}=\dot{M}_1l_1+\dot{M}_2l_2$, where $l_1=q^2/(1+q)^2$ and $l_2=1/(1+q)^2$ are the specific angular momenta of the binary components. 
When this is the case (only when, in addition to the accretion flow becoming circular near the accreting masses, the entire angular momentum accreted contributes to the orbital angular momentum of the system rather than the spin), Equation (\ref{eq:evolEqn1}) reduces to 
\begin{equation}\label{eq:dadtg}
\frac{d\log{a_b}}{d\log{M}} = \frac{2\dot{J}_g}{\dot{M}l_b} - 1.
\end{equation}
Notably, \textit{if} the aforementioned assumptions hold, the evolution of the binary semi-major axis only depends on the gravitational torque, overall accretion rate, and the specific angular momentum of the binary. 
The evolution of the mass ratio through accretion can be calculated similarly, using
\begin{equation}
\frac{d\log{q}}{d\log{M}}=\frac{1+q}{q}\frac{\dot{M}_2}{\dot{M}}-(1+q)\frac{\dot{M}_1}{\dot{M}}=\frac{1+q}{1+\lambda}\left(\frac{\lambda}{q}-1\right).
\end{equation}

We have also computed a number of one-dimensional radial histograms: binning in radius from both the binary barycenter and in coordinate systems centered on each objects, and also binning quantities as functions of fluid element semi-major axis about the barycenter. The semi-major axis of each fluid element is calculated from its radial location and the magnitude of its velocity ($v$) as $a=(2GM/r-v^2)^{-1}$, and the eccentricity vector of each fluid element is calculated as $\mathbf{e}=(v^2\mathbf{x}-(\mathbf{x}\cdot\mathbf{v})\mathbf{v})/GM - \mathbf{x}/r,$ and where $r\equiv|\mathbf{x}|$ is the distance from the binary barycenter.
Every time step, we include time- and either volume- or mass-weighted samples of fluid surface density, eccentricity, and gravitational torque density. We use the lattermost to calculate radially integrated gravitational torque profiles $T_g\equiv-\int_0^rdr\int_0^{2\pi}\Sigma(\mathbf{x}\times\nabla\Phi)d\phi$. 

We used these histograms to measure various quantities describing the disk. For example, we use histograms of the surface density and mass-weighted eccentricity, as functions of binary semi-major axis, to determine quantities such as the characteristic cavity semi-major axis ($a_{\rm cav}$) and eccentricity ($e_{\rm cav}$). We define the cavity location as that of the minimum of the time-averaged $|\partial^2\Sigma(a)/\partial a^2|$ for $a_b<a<6a_b$, which is valid for all of our simulations, as the surface density profile $\Sigma(a)$ always has an inflection point at $a>a_b$. We define the cavity eccentricity as the average eccentricity of the disk at $a_{\rm cav}$. We derived radial profiles of the disk argument of periapsis over time using the mass-weighted histograms of $\mathbf{e}$, defining the argument of periapsis as $\omega=\tan^{-1}{e_y/e_x}$. We measured the average disk precession rate by measuring the time derivative of $\omega$ between $a=a_{\rm cav}$ and $a=a_{\rm cav}+a_b$.

\subsection{Initial Conditions and Discretisation}\label{sec:setup}
The domain of our simulations was centered on the binary barycenter. We initialized each simulation with an axisymmetric surface density distribution of the form
\begin{equation}\label{eq:initSig}
\Sigma(r,t=0)=\Sigma_0\left[\eta + (1-\eta)e^{\left(-\left(\frac{r}{r_c}\right)^{-\xi}\right)}(1-\ell_0r^{-1/2})r^p\right],
\end{equation}
where $r_c$ sets the initial radius of a cavity around the binary, $\xi$ sets the steepness of this cavity and $\eta$ is a constant which sets the minimum density in our initial condition, $p$ sets the surface density profile far from the binary and is given by $p=0$ for simulations using a constant kinematic viscosity and  $p=-1/2$ for simulations using an $\alpha$-viscosity prescription, $\ell_0$ is an initial guess for the eventual values of $l_0$ in each simulation, and $\Sigma_0$ is a fiducial surface density. Setting $\ell_0$ near $l_0$ has been shown to significantly expedite the process of simulations settling into steady states of globally constant (in a time-averaged sense) accretion rate \citep{2017MNRAS.466.1170M} and prevent temporary suppression of the accretion rate onto the binary at high Mach numbers \citep{2022MNRAS.513.6158D}.
We set $r_c=2.5a_b$, $\xi=10$, and $\eta=0.0001$ in all of our simulations.  The initial radial velocity in our  simulations was given by
\begin{equation}
v_r(r,t=0) = -\frac{3\nu}{2\mathcal{R}},
\end{equation}
where $\mathcal{R}=\sqrt{r^2+\epsilon_g^2}$. The initial azimuthal velocity ($v_\phi=r\Omega$) is specified through approximate hydrostatic equilibrium taking into account both pressure gradients and the binary quadrupole moment, 
\begin{equation}
\Omega^2(r,t=0)=\Omega_k^2(\mathcal{R})\left(1+\frac{3a_b^2}{\mathcal{R}^2}\frac{q}{(1+q)^2}\right)+\frac{1}{\mathcal{R}\Sigma}\frac{d\Pi}{dr}
\end{equation}

Our simulations are carried out in Cartesian coordinates, their domains extending from $-10a_b$ to $10 a_b$ in both $x$ and $y$. Although we used outflow boundary conditions, we also set the source terms $D_\Sigma$ and $\mathbf{D}_p$ to damp fluid variable towards their initial values near the outer boundary. For each conserved variable $U_i$, the corresponding source term $D_i$ takes the form
\begin{equation}
D_i = (U_i(t=0) - U_i)f(r)\beta\frac{v_\phi}{r},
\end{equation}
where 
\begin{equation}
f(r)=\left\{ 
\begin{array}{ll}
0 & r<r_d \\
1 & r>r_d + l_d \\
1-(1-((r-r_d)/l_d)^2)^2 & \rm otherwise, \\
\end{array}
\right.
\end{equation}
and we set $\beta=10,$ $r_d=9.5a_b$, and $l_d=0.5a_b$. These damping terms serve to suppress $m=4$ instabilities related to our use of a Cartesian grid, and have been employed previously in \citet{2023ApJ...949L..30D}.

Our simulations at high mass ratios ($q\geq0.1$) utilized static mesh refinement, refining the grid by a factor of two within $[-5a_b,5a_b],$ $[-2.5a_b,2.5a_b],$ and $[-1.25a_b,1.25a_b]$. We typically resolve our base grid using $384$ cells along each dimension, reaching a resolution of $\Delta x = \Delta y \approx 0.0065a_b$ at the highest refinement level. These simulations held $\epsilon_{g,i}=0.035a_b$ and $r_{s,i}=0.035 a_b$. For our $\nu=0.001\Omega_ba_b^2$ simulation we set $\gamma_i=2.67$, and for our $\nu=0.0005\Omega_ba_b^2$ simulations and those using an $\alpha-$viscosity prescription we set $\gamma_i=1.33$. 

\begin{figure*}
\includegraphics[width=\linewidth]{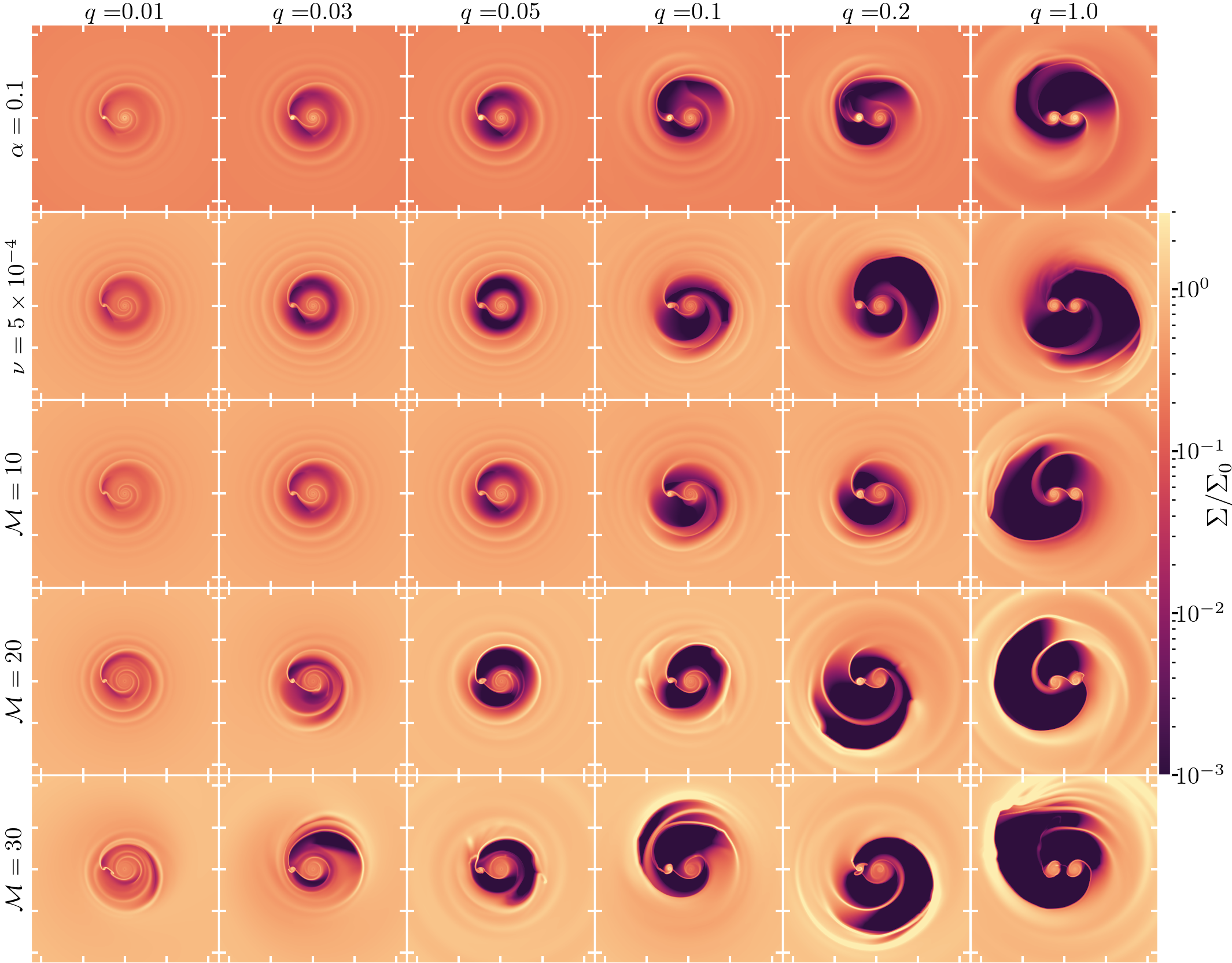}
\caption{Surface density profiles from a subset of our simulations, increasing in binary mass ratio from left to right. Profiles from simulations with $q<0.1$ are presented at $t=1000\pi$, and profiles from simulations with $q\geq0.1$ are presented at $t=4000\pi$.  
Snapshots from simulations with $q\geq0.1$. Results from simulations of $\mathcal{M}=10$ disks with $\alpha=0.1$ and $\nu=0.0005a_b^2\Omega_b$ are presented in the first and second rows respectively. The third, fourth, and fifth rows display results from simulations of $\nu=0.001a_b^2\Omega_b$ disks at $\mathcal{M}=\{10,20,30\}$ respectively. In all cases deeper and larger cavities form at higher mass ratios. At higher Mach numbers the cavity remains visibly eccentric even at very low mass ratios, although in thicker disks the cavity eccentricities driven by lower mass ratio binaries become more subtle. Tick markings are spaced $2a_b$ apart.}
\label{fig:gridBig}
\end{figure*}

For our simulations at low mass ratios ($q<0.1$), we utilize the adaptive mesh refinement framework (although in a static manner) to keep the largest cell size within each mesh block as close to $\Delta x \approx \rm{max}[(r-r_2),r_c]\epsilon$ as possible, where $\epsilon$ is a target cell aspect ratio, $r_c$ is a cutoff length, and $(r-r_2)$ is the difference between the radial location of a given cell and the radius of the orbit of the secondary. This choice of refinement focuses resolution near the orbit of the secondary while not requiring very small cells near the much more massive primary, allowing us to resolve the flow of gas within the Hill sphere of the secondary (of radius $r_H= a_b(q/3)^{1/3}$) while reducing the time step penalties incurred by uniformly resolving the inner regions of the accretion flow. We typically set $\epsilon=4/300$ and $r_c=0.15a_b$ over a grid with a base resolution of $\Delta x\approx$, resulting in cell sizes of $\Delta x \approx 0.0016 a_b$ near the secondary and $\Delta x \approx 0.0065 a_b$ near the primary. 

For our low-mass ratio runs, we scale the softening and sink radii of each point mass according to $\epsilon_{g,i}/\epsilon_{g,0}=r_{s,i}/r_{s,0}=g(m_i)$, where $g(m_i)=(2m_i)^{1/3}$ so that the sink and softening lengths are scaled according to the tidal radius about each particle, and set $r_{s,0}=\epsilon_{g,0}=0.035a_b$. We scale the sink rate according to $\gamma_i=\gamma_0g(m_i)^2$ in our constant-$\nu$ simulations, and according to $\gamma_i=\gamma_0g(m_i)^{3/2}$ for constant-$\alpha$ simulations to keep a roughly constant ratio of the viscous timescale at the sink radius $\sim r_{s,i}2/\nu$ to the accretion timescale ($\sim\Omega_b^{-1}/\gamma$). We set $\gamma_i=2.67$ in our $\nu=0.001\Omega_ba_b^2$ simulations, and for our $\nu=0.0005\Omega_ba_b^2$ simulations and those using an $\alpha-$viscosity prescription we set $\gamma_i=1.33$. 

All simulations used $3rd-$order piecewise-parabolic reconstruction \citep{2018JCoPh.375.1365F}, the second-order van Leer integrator \citep{2009NewA...14..139S} with a CFL factor of 0.15, and the Roe approximate Riemann solver \citep{1981JCoPh..43..357R}. We have found that, particularly at low mass ratios, high-order spatial reconstruction schemes significantly aid convergence. We chose initial values of $\ell_0$ for each Mach number based on the simulations presented in \citet{2022MNRAS.513.6158D} and the observation that the angular momentum current through the disk varies quite weakly with binary mass ratio, at least at $\mathcal{M}=10$ \citep[e.g.,][]{2020ApJ...889..114M,2021ApJ...921...71D}, using $\ell_0=\{0.7,0.1,-0.5\}$ for $\mathcal{M}=\{10, 20, 30\}$ respectively.

\section{Results} \label{sec:results}

\begin{figure}
\includegraphics[width=\linewidth]{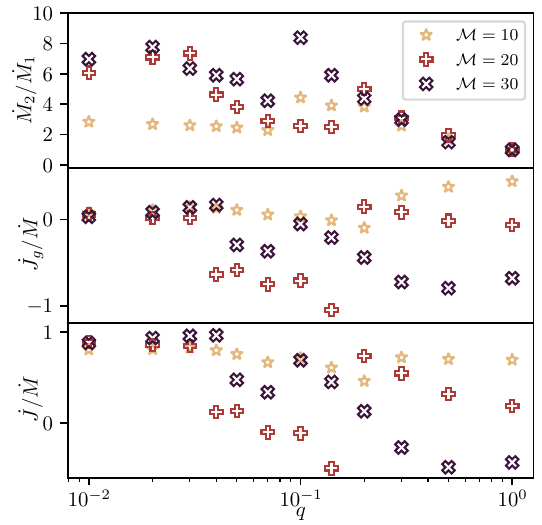}
\caption{Values of $\dot{M}_2/\dot{M}_1$, $\dot{J}_g/\dot{M}$, and $\dot{J}/\dot{M}$ from $q=0.01$ to $q=1$ at $\mathcal{M}\in\{10,20,30\}$ disks with $\nu=0.001$, averaged in time over the last 500 orbits of each simulation. Although all quantities are non-monotonic, as soon as symmetry is broken and $M_2<M_1$, $\dot{M}_2>\dot{M}_1$. Although, when moving from high to low mass ratios, gravitational torques initially grow weaker, the gravitational torque suddenly (in $q$) becomes negative, at mass ratios which depend on Mach number. These negative torques correspond to decreases in $\dot{M}_2/\dot{M}_1$ in the $\mathcal{M}=20$ and $\mathcal{M}=30$ cases, which are discussed in Section \ref{sec:nomode}. Numerical values are provided in Table \ref{tab:summary2}.}
\label{fig:machsurvey}
\end{figure}

\begin{figure}
\includegraphics[width=\linewidth]{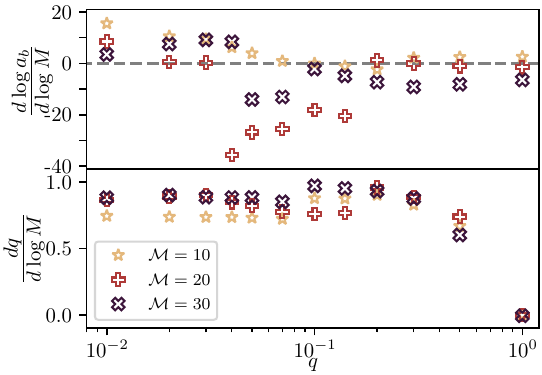}
\caption{The evolution of binary mass ratios and semi-major axes in the same simulations shown in Figure \ref{fig:machsurvey}. At larger mass ratios, we find results in general agreement with previous studies. However, at intermediate mass ratios, which have not been previously probed at $\mathcal{M}\neq10$, we find extremely rapid inspirals, following from both the appreciably negative torque, small specific angular momentum of the binary, and that $\lambda>q$. At very low mass ratios ($q\lesssim0.03$), the rate of change of the binary semi-major axis can be quite large unless the torque is very nearly balanced by the change in the mass of the binary. Numerical values are provided in Table \ref{tab:summary2}.}
\label{fig:machsurveyLogs}
\end{figure}

\begin{figure}
\includegraphics[width=\linewidth]{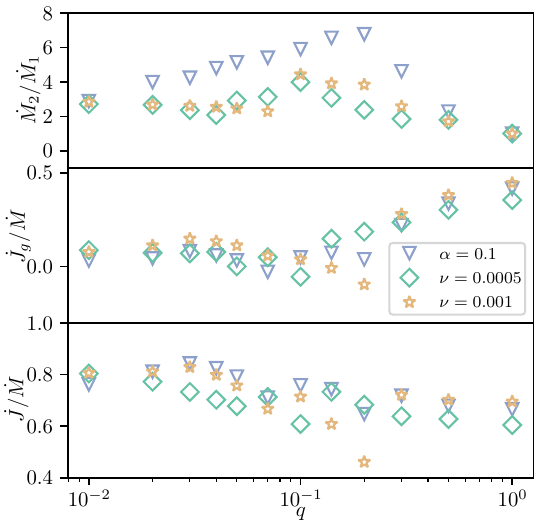}
\caption{Values of $\dot{M}_2/\dot{M}_1$, $\dot{J}_g/\dot{M}$, and $\dot{J}/\dot{M}$ from $q=0.01$ to $q=1$ with $\mathcal{M}=10$ and disks with viscosities $\nu=0.001,$ $\nu=0.0005$, and $\alpha=0.1$, averaged in time over the last 500 orbits of each simulation. At all mass ratios below $1$, we find that binaries accreting from $\alpha-$viscous disks accrete more preferentially onto the secondary than those accreting from constant$-\nu$ disks. Depending on the viscosity prescription, binaries at intermediate mass ratios, $0.05\lesssim q \lesssim 0.2$ experience negative gravitational torques, which we discuss further in Section \ref{sec:thickecc}. Numerical values are provided in Table \ref{tab:summary2}.}
\label{fig:viscsurvey}
\end{figure}
\begin{figure}
\includegraphics[width=\linewidth]{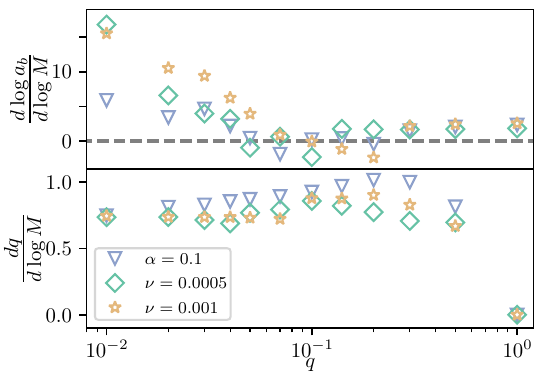}
\caption{The evolution of binary mass ratios and semi-major axes in the same simulations shown in Figure \ref{fig:viscsurvey}. We find that all unequal-mass binaries are driven towards $q=1$, although binaries accreting from $\alpha$ disks accrete towards equal mass at a faster rate. At intermediate mass ratios, where the gravitational torque is negative or small in magnitude (see Equation \ref{eq:dadtg}), binaries are driven to inspiral, while at other mass ratios the binaries can outspiral rather quickly. Notably, the rate of outspiral at low mass ratios varies strongly with viscosity, disks with constant kinematic viscosity leading to significantly faster outspirals. Numerical values are provided in Table \ref{tab:summary2}.}
\label{fig:viscsurveyLogs}
\end{figure}

To begin a sweeping overview of our simulations, we present snapshots of the gas surface densities from a subset thereof in Figure \ref{fig:gridBig}; a number of trends are immediately apparent. Across all disk viscosities and Mach numbers, higher mass ratio binaries open deeper and more pronounced cavities,\footnote{In the context of protoplanetary disks, the literature often discusses ``gaps,'' while in the literature on circumbinary disks frequently discusses ``cavities.'' We have adopted the latter throughout, but the former may also be appropriate to the gas distribution around some of the lowest-mass ratio binaries with $\mathcal{M}=10$ disks.} which follow directly from the weaker shocks driven by commensurately weaker torques from lower mass ratio binaries. 
Higher mass ratio binaries often drive more eccentric cavities, but this is not always the case: for example, out of our $\mathcal{M}=20$ simulations, the disk orbiting the $q=0.03$ binary is more eccentric than that around the $q=0.05$ binary. Previous studies of equal-mass binaries have noted that inner edges of higher Mach number circumbinary disks have higher surface densities than those around lower Mach number disks \citep[e.g.,][]{2020ApJ...900...43T}, which relates to the angular momentum current through those disks as demonstrated in \citet{2022MNRAS.513.6158D}; we observe here the same trend at lower mass ratios, at least for $q\gtrsim0.1$. Additionally, as noted by previous studies of protoplanetary and circumbinary disks, lower viscosity disks are characterized by deeper and wider cavities at a given mass ratio \citep[e.g.,][]{2018ApJ...861..140K,2022MNRAS.513.6158D}. 

\begin{figure}
\includegraphics[width=\linewidth]{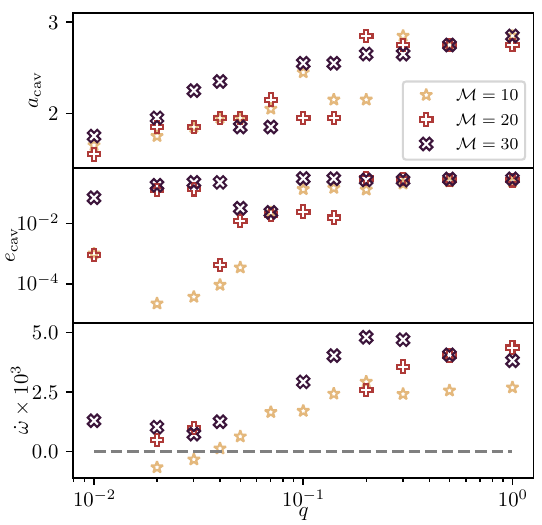}
\caption{Orbital properties of the inner edge of the circumbinary disk cavity/gap - the semi-major axis, scalar eccentricity, and argument of periapsis - from $q=0.01$ to $q=1$ at $\mathcal{M}\in\{10,20,30\}$ disks with $\nu=0.001$, averaged in time over the last 500 orbits of each simulation. Although all three quantities tend to decline as mass ratios decrease from $q=1$, they do not so monotonically. Notably, at certain intermediate mass ratios at $\mathcal{M}=20$ and $\mathcal{M}=30$, the cavity becomes much smaller, more circular, and ceases to precess in a meaningful way; however, even in these cases the typical magnitude of the disk eccentricity is larger than for some lower-$q$ binaries which still have well-defined, precessing, eccentric disks. Interestingly, at low mass ratios, some $\mathcal{M}=10$ disks precess in a retrograde rather than prograde sense. Numerical values are provided in Table \ref{tab:summary2}.}
\label{fig:machsurveyDisk}
\end{figure}

\begin{figure}
\includegraphics[width=\linewidth]{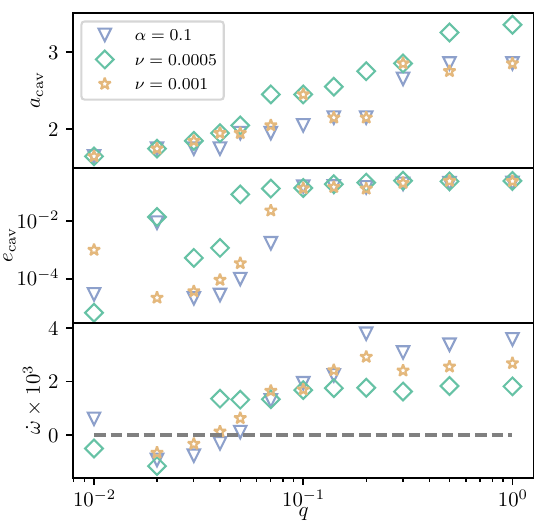}
\caption{The same disk properties plotted in Figure \ref{fig:machsurveyDisk}, but for our simulations of $\mathcal{M}=10$disks with viscosities $\nu=0.001,$ $\nu=0.0005$, and $\alpha=0.1$. The lowest constant-viscosity disks, at least for $q\gtrsim0.1$, tend to have larger cavities and precess more slowly. However, at those same higher mass ratios disks with $\alpha=0.1$ tend precess appreciably faster than disks with $\nu=0.001$, even when they have smaller cavities, illustrating the important dynamical differences which arise from different spatial viscosity profiles. Generally, as binary mass ratio decreases, so does the disk precession rate, albeit non-monotonically. Similarly, at $q\lesssim0.05$, many disks exhibit retrograde precession, although curiously the $\alpha=0.1$ disk around a $q=0.01$ binary precesses in a prograde sense. Numerical values are provided in Table \ref{tab:summary2}.}
\label{fig:viscsurveyDisk}
\end{figure}

We provide visual summaries of the quantities related to binary orbital evolution ($\dot{M}_2/\dot{M}_1,~\dot{J}_g/\dot{M},~\dot{J}/\dot{M},~d\log{a}_b/d\log{M}$, and $dq/d\log{M}$) as functions of mass ratio in Figures \ref{fig:machsurvey} and \ref{fig:machsurveyLogs}, for simulations which varied $\mathcal{M}\in\{10,20,30\}$ at constant kinematic viscosity $\nu=0.001$, and in Figures \ref{fig:viscsurvey} and \ref{fig:viscsurveyLogs} for simulations of $\mathcal{M}=10$ disks that employed Navier-Stokes viscosities of different magnitudes and spatial variations. Figures \ref{fig:machsurveyDisk} and \ref{fig:viscsurveyDisk} plot the similarly partitioned cavity semi-major axis, eccentricity, and precession rate. Numerical values are collected in Appendix \ref{app:summary}.

Only a few trends appear to be universal. The total torque on binaries in $\mathcal{M}=10$ disks is always positive (although in some cases the gravitational torque on the binary is negative). Additionally, unequal-mass binaries always accrete such that their mass ratio increases over time, and that disks with $\alpha$ viscosities accrete more preferentially onto the secondary than those with globally constant kinematic viscositites. 

The semi-major axis and eccentricity of the inner edge of the circumbinary disk tend to increase as the binary mass ratio increases, although not monotonically. At higher mass ratios the circumbinary disk always precesses in a prograde sense, and a rate which decreases (non-monotonically) at lower mass ratios. For the $\mathcal{M}=10$ disks considered here, the precession rate becomes negative, or retrograde with respect to the binary, at many (but not all) mass ratios below $q\lesssim0.05$.

At mass ratios between $q\sim0.1$ and $q\sim1$, a few other trends trends emerge across disk parameters. For example, material tends to accrete more and more preferentially onto the secondary as mass ratios decrease from $q=1$, $\dot{M}_2/\dot{M}_1$ peaking, at least locally in $q$, between $q=0.1$ and $q=0.2$. The gravitational torque on the binary also becomes gradually weaker, lessening in magnitude - this can be qualitatively understood as the accretion flow near the more massive primary approaching axisymmetry at smaller mass ratios, while the secondary shrinks in mass and gravitational influence. We examine these trends more quantitatively in the following subsections, before moving on to features more localized in mass ratio such as the sudden decreases in the gravitational torque at in $\mathcal{M}=30$ disks at mass ratios $q=0.05$ and $q=0.07$. We find appreciable changes in neither binary orbital evolution nor disk morphology related to the stability of particle orbits about the fourth and firth Lagrange points of the circular restricted three body problem, contrary to the advocation of \citet{2016MNRAS.459.2379D} - evidently viscous and hydrodynamical effects are more significant in this regime. 

\subsection{Accretion morphology around $q\gtrsim0.1$ binaries}
We begin by reviewing simulations which display characteristics most similar to those of equal-mass binaries, which have been studied in great detail previously. Higher mass ratio binaries open pronounced eccentric cavities in their circumbinary disks. Whenever one of the binary members approaches the pericenter of the cavity, some material is captured by the binary while other gas remains unbound. The latter receives a gravitational `kick' and is launched across the cavity, as illustrated in Figure \ref{fig:gridBig}. 
As gas from the edge of the cavity is perturbed by the binary, material which leads the binary orbit causes a positive gravitational torque on the binary, and material which lags behind the binary orbit exerts a negative torque. The total torque on the binary is then derived from these gravitational torques and the angular momentum carried by gas as it is captured by the binary, as virtually all gas remains within the sphere of influence of the binary after entering \citep[e.g.][]{2022ApJ...932...24T}. This advected angular momentum then manifests as gravitational torques proximate to the binary and finally as an accretion torque. In $\mathcal{M}=10$ disks, the overall gravitational torque on the binary is usually positive, but in higher Mach number disks, proportionally more gas is launched through the cavity each pericenter passage than is accreted, leading to stronger negative torques on the binary in colder, thinner disks \citep{2022MNRAS.513.6158D}. 

\begin{figure}
\includegraphics[width=\linewidth]{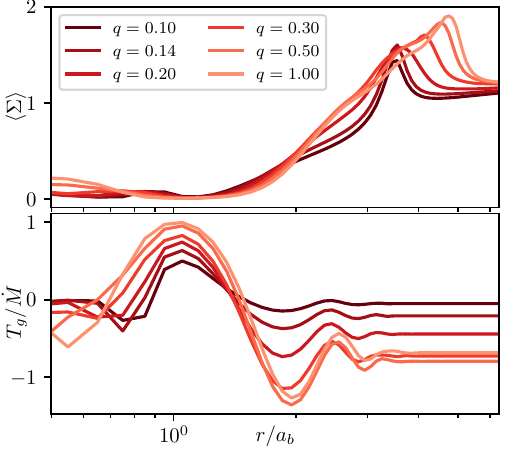}
\caption{Average surface density (top panel) and cumulative gravitational torque acting on the binary (bottom panel, the gravitational torque density integrated from $r'=0$ to $r'=r$, normalized by the accretion rate onto the binary) profiles corresponding to binaries with $0.1\leq q \leq 1$ accreting from $\mathcal{M}=30$ disks. Each profile was averaged over the final 500 binary orbits of each simulation.}
\label{fig:m30profs}
\end{figure}

\begin{figure}
\includegraphics[width=\linewidth]{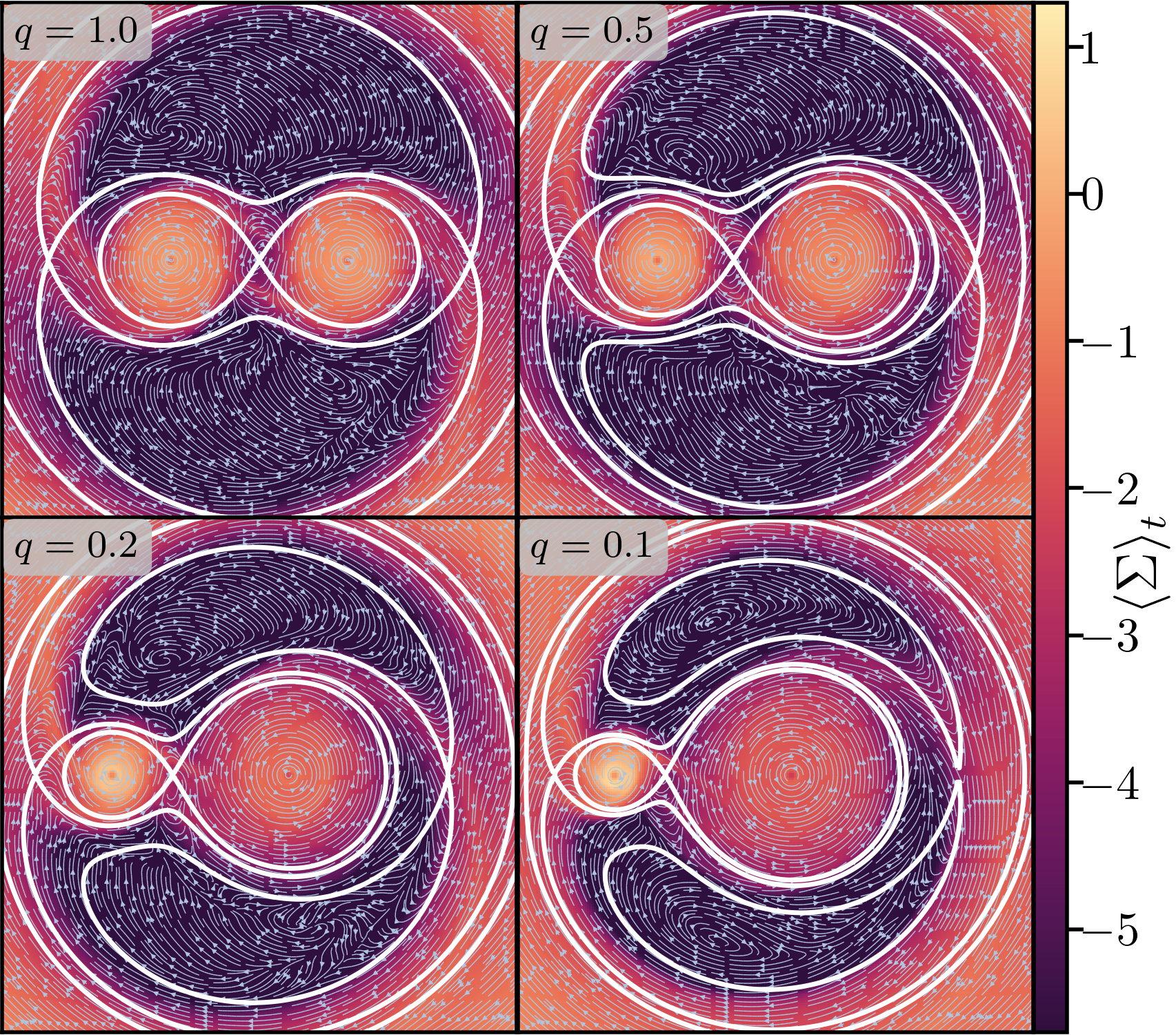}
\caption{Time-averaged, in the rotating frame of the binary, surface density profiles with momenta streamlines, for binaries with $\mathcal{M}=30$ disks at $q\in\{0.1, 0.2, 0.5, 1.0\}$. The equipotential contours of the first, second, and third Lagrange points are illustrated in each case using white curves. In a time-averaged sense, the flow onto the primary from the circumbinary disk transitions from both lagging and leading the primary at high mass ratios, to primarily trailing the binary at lower mass ratios, contributing to a less negative overall gravitational torque as $q$ decreases.}
\label{fig:sigmaM30}
\end{figure}
\begin{figure*}
\includegraphics[width=\linewidth]{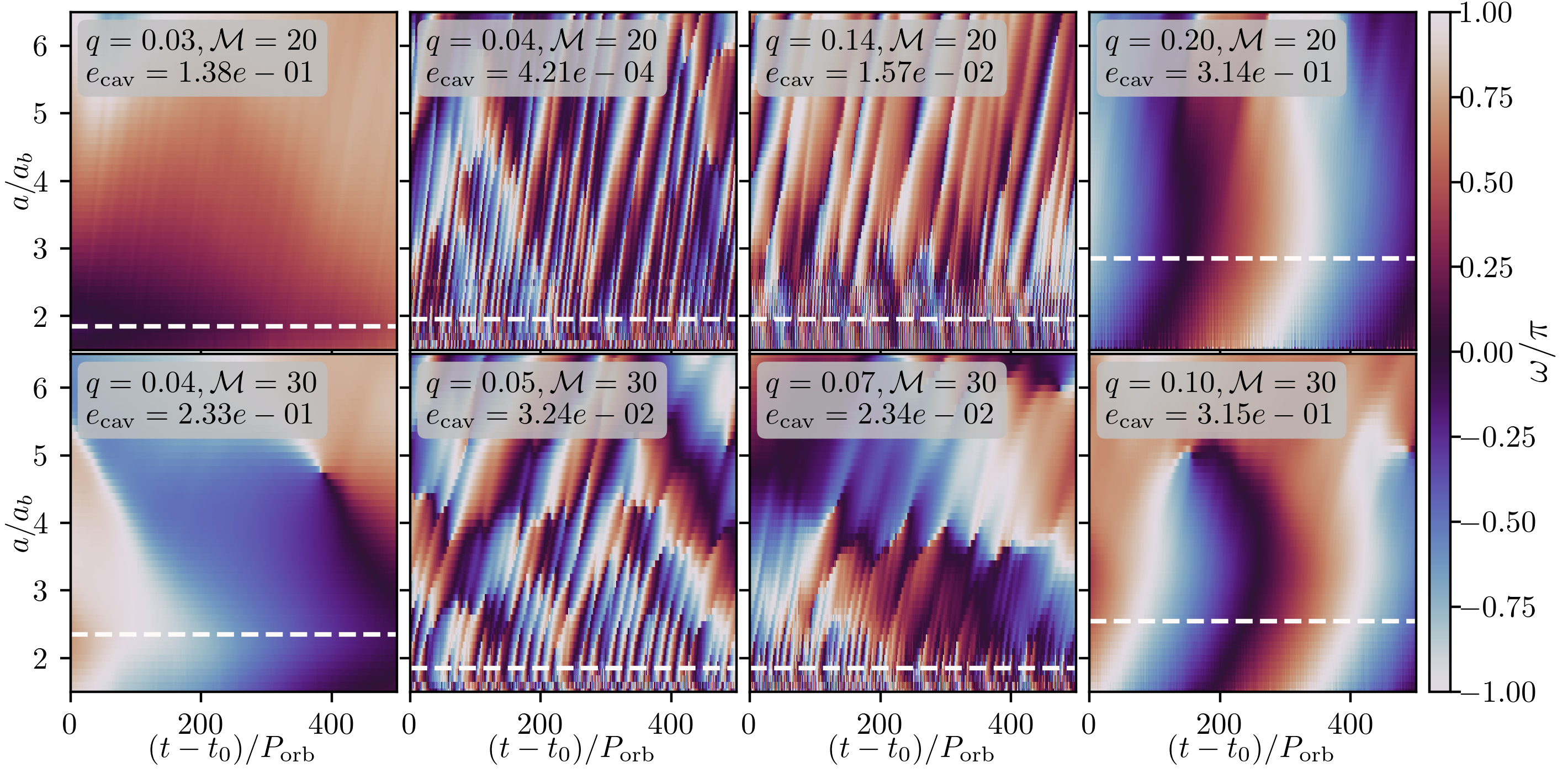}
\caption{Space-time diagrams of the disk argument of periapsis over the final 500 orbits of each simulation, averaged azimuthally in bins of constant fluid element semi-major axis, with the cavity semi-major axis marked using a dashed white line. Rarely is the entire disk precisely in phase. However, for each Mach number shown here, $\mathcal{M}=20$ in the top row and $\mathcal{M}=30$ in the bottom row, disks around both higher- and lower-mass binaries precess at similar rates throughout. However, disks around binaries at the intermediate mass ratios shown here lack a uniformly precessing eccentric mode, and their eccentricity at small semi-major axes is dominated by orbital-timescale variability. $P_{\rm orb}$ is the binary orbital period $2\pi\Omega_b^{-1}$. We used $t_0=1500\,P_{\rm orb}$ for simulations of binaries with $q\geq0.1$ and $t_0=500\,P_{\rm orb}$ for simulations with $q<0.1$.}
\label{fig:precTS}
\end{figure*}

We illustrate in Figure \ref{fig:m30profs} azimuthally averaged surface density and radially integrated torque profiles from our higher-mass ratio $\mathcal{M}=30$ simulations. To assist with interpretation, we also present time-averaged surface density profiles for a subset of these simulations, along with streamlines of the time-averaged momentum profiles, in the frame rotating with the binary, in Figure \ref{fig:sigmaM30}. Considering first the equal-mass binary, we note the net negative gravitational torque from $r\lesssim a_b/2$ shown in the bottom panel of Figure \ref{fig:m30profs}. This corresponds to the streams of gas which flow from one member of the binary to the other, illustrated in the top left panel of Figure \ref{fig:sigmaM30}, which lag behind the binary and exert a negative torque. From $r\sim a_b/2$ to $r\sim a_b$, a strong positive torque is exerted on the binary by gravitational interactions between each binary component and the opposite minidisk, as well as the portions of accreting gas streams which lead the binary. The final appreciable component of the gravitational torque then arises between $r\sim a_b$ and $r\sim 2a_b$, from gas which on average lags the binary, that which is stripped from the cavity walls only to be flung through the cavity rather than captured by the binary. 

As binary mass ratios decrease, the accretion morphology becomes more axisymmetric, many of the torque components weaken, and the circumbinary disk gradually shifts inward. As the secondary moves outward, so does the radius at which mass typically flows between primary and secondary minidisks, as illustrated in Figure \ref{fig:sigmaM30}. This shift corresponds to the low-radius region of negative torques extending to larger radii, but becoming less prominent, as illustrated in Figure \ref{fig:m30profs}. 
Pronounced positive torques arise both in the minidisk about the secondary and from gas accreting from the circumbinary disk at some radii. 
Material which accretes directly onto the primary tends to lead it, exerting a positive torque, while material which accretes onto the secondary leads the binary at smaller radii, but lags the binary at larger radii. At lower mass ratios, less material is perturbed by the secondary, and lower average surface densities within the cavity correspond to smaller torques. 
Similar trends occur for all disks around higher mass ratio binaries. 

\subsection{Cavity size, eccentricity, and gravitational torques}
\subsubsection{Thin disks and the absence of eccentric modes}\label{sec:nomode}

At intermediate mass ratios, binaries accreting from $\mathcal{M}=20$ and $\mathcal{M}=30$ disks experience strikingly non-monotonic gravitational torques at the same mass ratios where the eccentricity of their circumbinary disks also drops dramatically. These transitions occur at $\mathcal{M}=20$ between $q=0.04$ and $q=0.14$ (inclusive), and in $\mathcal{M}=30$ disks only at $q=0.05$ and $q=0.07$. As illustrated in Figure \ref{fig:precTS} and the precession rates and cavity eccentricities plotted in Figures \ref{fig:machsurveyDisk} and \ref{fig:viscsurveyDisk}, the circumbinary disks corresponding to these extreme negative torques are not simply near-circular, but lack a quasi-global, precessing, eccentric mode throughout. In particular, a number of the disks around $q=0.01$ binaries have even lower average eccentricities, but sill possess large-scale uniformly precessing eccentric modes. 

\begin{figure}
\includegraphics[width=\linewidth]{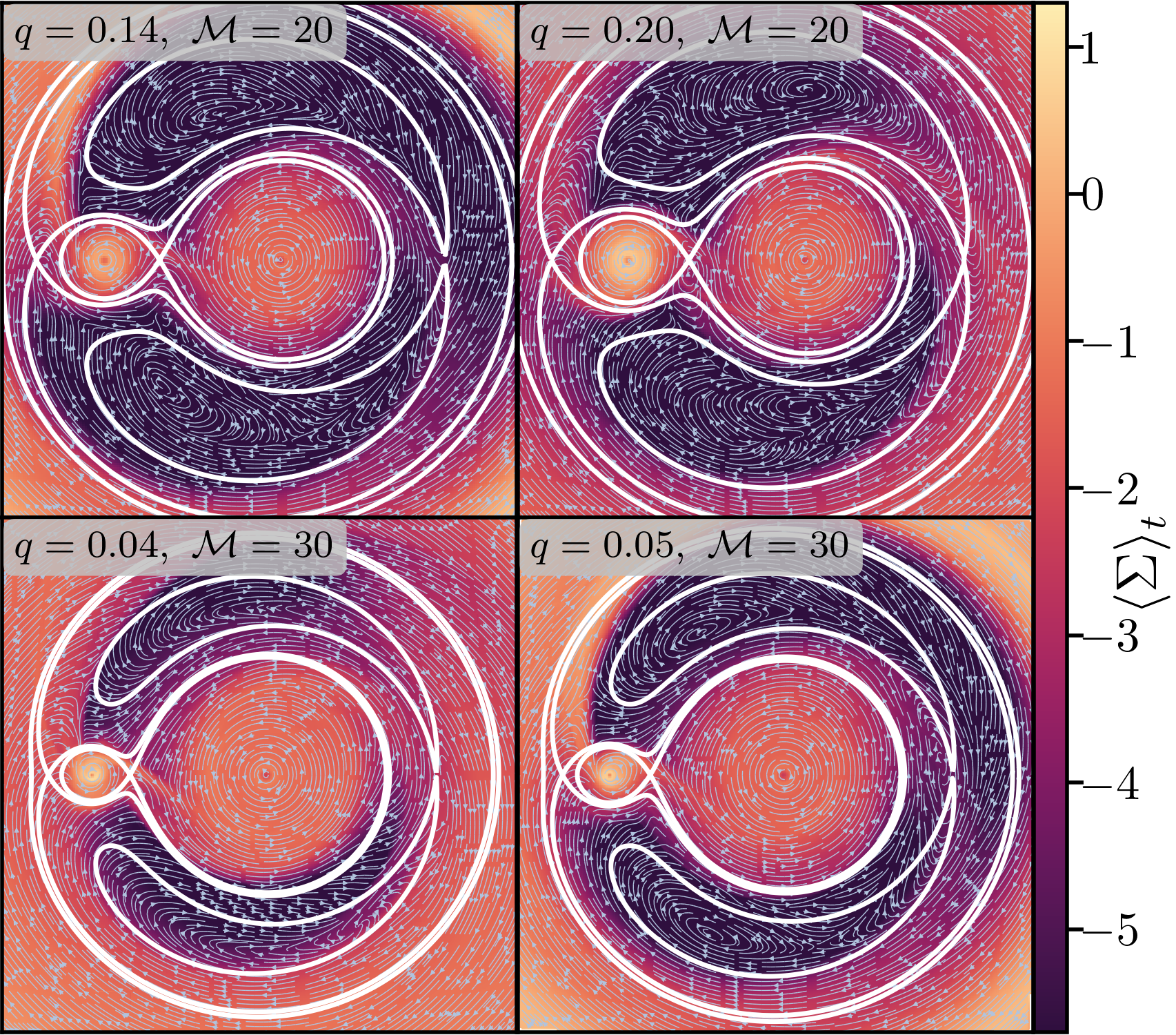}
\caption{Time-averaged profiles of the disk surface density, with streamlines of the time-averaged momentum field, in the frame of the binary for a few of the simulations depicted in Figure \ref{fig:precTS} in which eccentric modes appear at intermittent mass ratios. As binaries accrete from disks with prominent eccentric modes ($q=0.2~\mathcal{M}=20$ and $q=0.04~\mathcal{M}=30$ here), gas that flows onto the primary more often leads it, contributing to a more positive torque. Additionally, disks with eccentric modes accrete in a more spatially uniform manner onto the secondary, while the secondary-bound accretion streams from disks without eccentric modes tend to lag behind the binary. The aforementioned effects conspire to create a much more negative gravitational torque on binaries accreting from disks; the difference in torque follows from a fundamental change in the accretion flow morphology, and the primaries actually accrete (proportionally) more from disks without eccentric modes than from disks with prominent eccentric modes.}
\label{fig:sigmaEccAvg}
\end{figure}

The low disk eccentricities of and strong negative gravitational torques upon binaries at these mass ratios are illustrated in Figures \ref{fig:machsurveyDisk} and \ref{fig:machsurvey} respectively. The absence of coherent eccentric modes in these systems, and the presence of such modes at both higher and lower mass ratios, is illustrated in Figure \ref{fig:precTS}, which plots the azimuthally averaged disk argument of periapsis as a function of space and time for a set of binaries. The physics governing the growth, saturation, and precession rate of eccentric modes is beyond the scope of this work, but we plan to pursue a deeper understanding thereof in a future study. However, we observe steady precession, on periods of hundreds to thousands of binary orbits, in disks which support eccentric modes, and a lack of such precession in those which do not. Notably, as cavity semi-major axes and eccentricities drop as eccentric modes disappear at certain mass ratios, the pericenter of the cavity stays nearly constant in most cases (\ref{tab:summary2}).

Some of the resulting changes to the accretion flow are illustrated in Figure \ref{fig:sigmaEccAvg}, which plots time-averaged surface density profiles for a pair of binaries with eccentric and non-eccentric disks at both $\mathcal{M}=20$ and $\mathcal{M}=30$. A few of the relevant binaries are also featured in Figure \ref{fig:gridBig}, particularly those at $q=0.03$ and $q=0.05$. Although not obvious from the time-averaged surface density profiles, circumbinary disks without eccentric modes at these mass ratios accrete much more onto the primary than those with eccentric modes. However, the morphology of this accretion changes dramatically, almost entirely lagging behind the primary rather than leading it and thus exerting a negative torque rather than positive. The streams of gas which accrete onto the secondary from disks without eccentric modes also appear to be higher-density, and more asymmetric azimuthally, lagging the binary more pronouncedly and further contributing to the strong negative torques in these systems. 

\subsubsection{Thick disks and mild torques}\label{sec:thickecc}
\begin{figure}
\includegraphics[width=\linewidth]{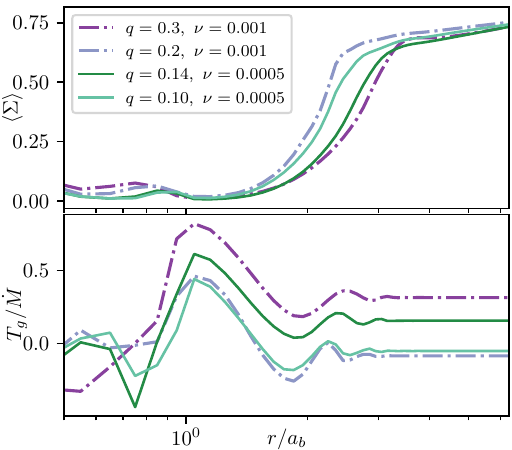}
\caption{Average surface density (top panel) and cumulative gravitational torque (bottom panel, the gravitational torque density integrated from $r'=0$ to $r'=r$, normalized by the accretion rate onto the binary) profiles for binaries accreting from $\mathcal{M}=10$ disks. Similar changes in both the surface density and cumulative torque profiles occur for both viscosities, the disk moving inwards and the overall torque becoming negative, although these changes  Each profile was averaged over the final 500 binary orbits of each simulation.}
\label{fig:profsVisc}
\end{figure}

Thicker, $\mathcal{M}=10$ disks also exhibit torques which become slightly negative at intermediate mass ratios, which are often accompanied by changes in cavity size and eccentricity (c.f. Figures \ref{fig:viscsurvey} and \ref{fig:viscsurveyDisk}). Notably, \citet{2020ApJ...889..114M} and \citet{2021ApJ...921...71D} observed, in $\alpha=0.1$ and $\nu=0.001$ disks respectively, smaller torques on binaries with $q=0.2$ than those with $q=0.1$ of $q=0.3$. We have reproduced these results herein, and found similar behavior in disks with $\nu=0.0005$ at $q=0.1$. While it is tantalizing to guess that the same mechanisms at play in the previously discussed $\mathcal{M}=20$ and $\mathcal{M}=30$ disks might be at play here, these $\mathcal{M}=10$ disks still exhibit uniformly precessing eccentric modes even when the magnitude of their eccentricity drops at these mass ratios. Additionally, while the cavity size and eccentricity change, they do not do so while keeping nearly constant preicenter as did the higher Mach number disks (Table \ref{tab:summary2}), and the secondary actually accretes more preferentially as the torque drops whereas more accretion onto the primary occurred from more circular disks in the higher Mach number cases. 

\begin{figure}
\includegraphics[width=\linewidth]{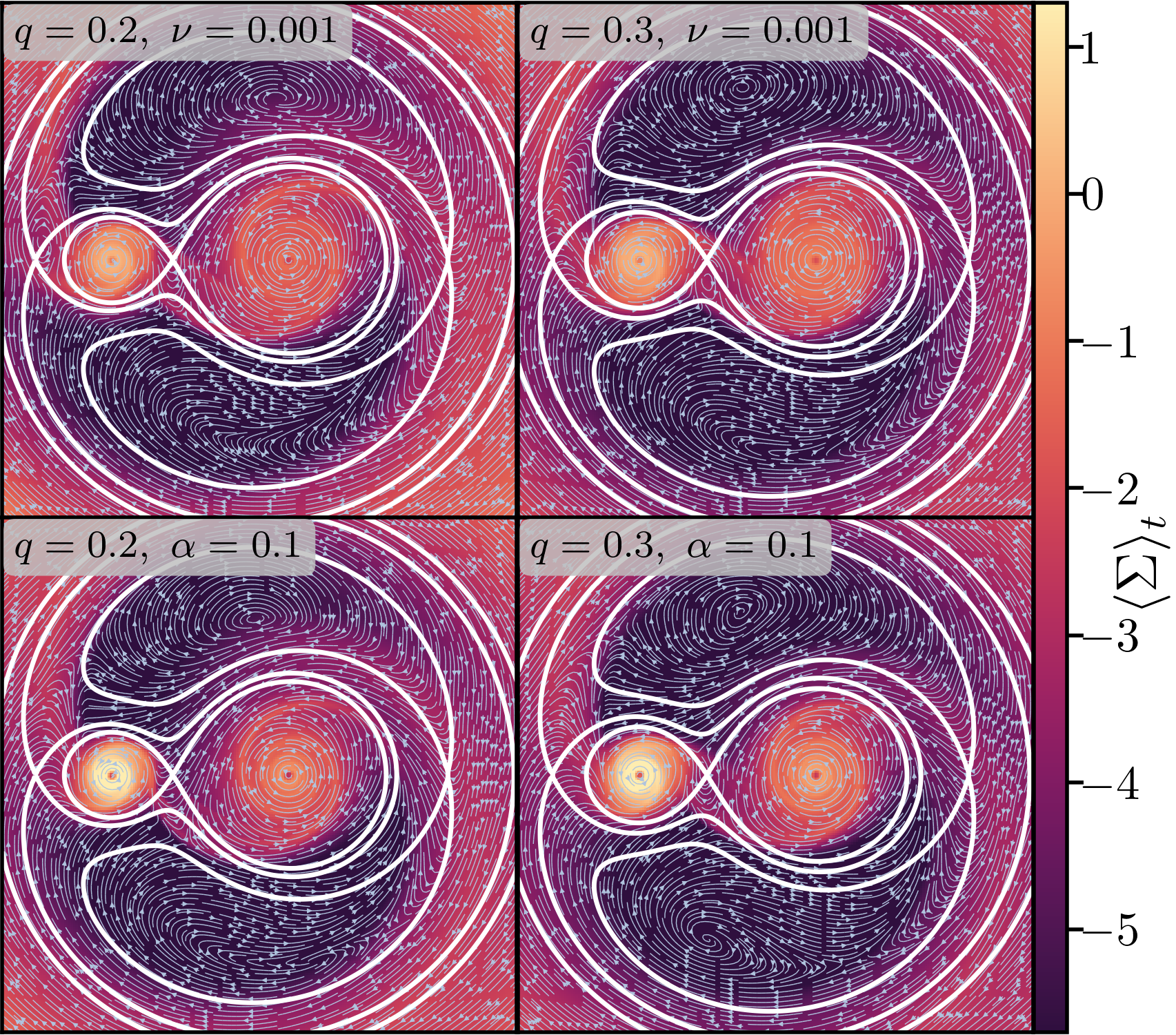}
\caption{Time-averaged profiles of the disk surface density, with streamlines of the time-averaged momentum field, in the frame of the binary for a few binaries accreting from $\mathcal{M}=10$ disks. At lower mass ratios more gas which is stripped from the cavity by the primary eventually becomes bound to the secondary, and less gas flows from the minidisk of the secondary onto the primary. These morphological changes coincide with decreases in the gravitational torques at $q=0.2$ for these systems.}
\label{fig:sigmaViscGrid}
\end{figure}

To illuminate some of the causes of these trends in gravitational torque, which occur at different mass ratios for different disk viscosities, we plot in Figure \ref{fig:profsVisc} a few profiles of surface density and integrated gravitational torque density for pertinent mass ratios. For the $\nu=0.001$ disk, the change in cavity size between the $q=0.3$ and $=0.2$ binaries is similar to that between the $q=0.14$ and $q=0.1$ binaries at $\nu=0.0005$, suggesting that the same mechanism governs both cases. The integrated torque profiles in all cases largely follow each other with nearly constant offsets beyond $r\gtrsim a_b$, suggesting that the torque contributions from larger radii, including the negative torques dominant from $a_b\lesssim r \lesssim 2a_b$, are largely similar for each mass ratio and viscosity. 
However, changes to the morphology at small radii occur according to the angular momentum carried by the infalling gas. Upon examination of time-averaged surface density profiles, shown in Figure \ref{fig:sigmaViscGrid}, it appears that as the mass ratio drops, an appreciable amount of gas which flows about the primary finds its way into the Roche lobe of the secondary, not remaining bound to the primary. Additionally, at the lower mass ratios in these cases, less gas seems to flow from the secondary to the primary, in accordance with the larger values of $\dot{M}_2/\dot{M}_1$ observed at the lower mass ratios of the binaries considered here. 

\subsection{Trends in preferential accretion}
At all non-unity mass ratios, we observe that the secondary accretes at a higher rate than the primary. While many previous studies have identified this trend at $\mathcal{M}=10$, recent studies have found various degrees of (dis-)agreement in the magnitude of $\dot{M}_2/\dot{M}_1$. Despite prior studies using constant-$\alpha$ viscosity prescriptions \citep[e.g.,][]{2020ApJ...889..114M,2022arXiv220302514S}, constant-$\nu$ viscosity prescriptions \citep[e.g.,][]{2021ApJ...921...71D}, and heterodox formulations of the equations of viscous hydrodynamics \citep{2020ApJ...901...25D}, the role of viscosity as been heretofore unappreciated: $\alpha$-viscosity, at least in $\mathcal{M}=10$ disks, enhances the degree of preferential accretion onto the secondary, as shown in Figure \ref{fig:viscsurvey}. Additionall, at very low mass ratios, we find a higher degree of preferential accretion in colder disks, as shown in Figure \ref{fig:machsurvey}.

\begin{figure*}
\includegraphics[width=\linewidth]{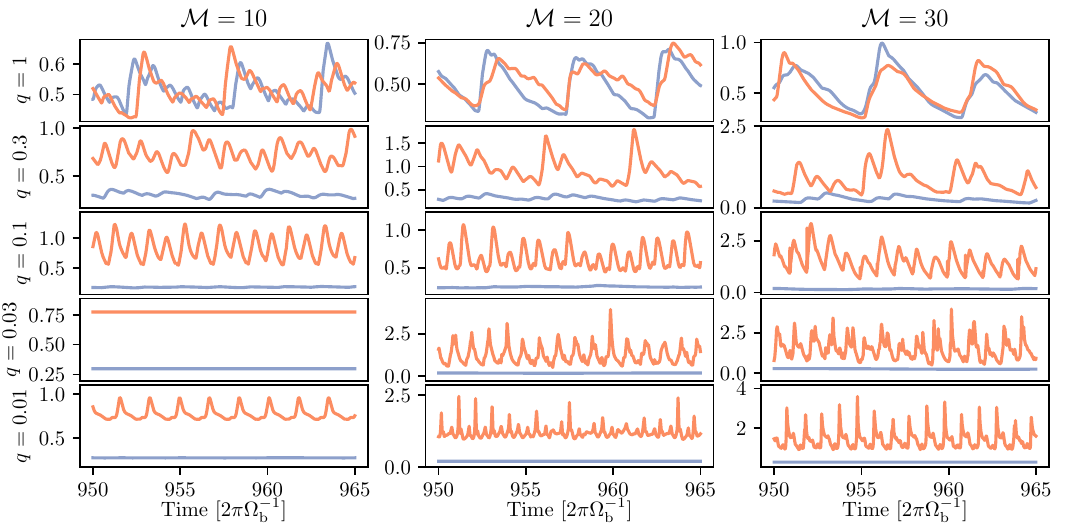}
\caption{Accretion rate time series, over fifteen orbital periods, for a representative sample of binaries of mass ratios. In each panel, the accretion rate onto the primary is plotted in blue ($\dot{M}_1/\dot{M}_0$), and the accretion rate onto the secondary is plotted in orange ($\dot{M}_2/\dot{M}_0$), in both cases normalized by the equilibrium accretion rate for each simulation. At lower mass ratios ($q\lesssim0.1$), the accretion rate onto the primary is fairly constant, and well below the accretion rate onto the secondary.}
\label{fig:timeseries}
\end{figure*}

\subsubsection{Viscosity}\label{sec:viscpref}
When equal partition of material between the primary and secondary is not guaranteed by symmetry, $\alpha$ viscosity causes higher rates of accretion onto the secondary relative to those onto the primary than in comparable simulations of disks with constant kinematic viscosity. Qualitatively, from examination and comparison of the constant-$\alpha$ and constant-$\nu$ binaries presented in Figure \ref{fig:sigmaViscGrid}, it appears that less matter flows between the primary and secondary in $\alpha-$viscous disks than in constant-$\nu$ disks, while the amount of gas flowing directly onto the primary through the cavity does not appear appreciably different. It may be that the relatively flat surface density profiles of constant-$\nu$ disks in Newtonian potentials allow more material to flow between the binary members, while the more centrally-peaked surface density profiles that arise from $\alpha$-viscosity prescriptions ($\Sigma\propto r^{-1/2}$ in a Newtonian potential) allow less gas to flow from the minidisk of the secondary to that of the primary. 

In an attempt to test this, we have carried out an additional simulation at $q=0.1,\mathcal{M}=10$ using an ad hoc viscosity prescription,\footnote{Although a different radial viscosity profile could be achieved within an $\alpha$-viscosity framework by varying the sound speed profile of the disk, such changes also alter gas dynamics even for constant viscosity profiles, as illustrated in \citet{2022MNRAS.513.6158D}, so we have restricted ourselves to a simpler scenario.} 
defining $\nu=\aleph c_s \text{\<.h>}$, setting $\aleph=0.1$ and defining
\begin{equation}
\text{\<.h>}=\frac{c_s\sqrt{a_b}}{\sqrt{\sum_i GM_i(|\textbf{x}-\textbf{x}_i|^2+\epsilon_g^2)^{-2}}}
\end{equation}
in analogy to Equation \ref{eq:alphavisc} but with a viscosity profile $\nu\propto r$ far from the binary and $\nu\propto r_i$ within each minidisk, resulting in a more sharply peaked surface density profile within the minidisks. We find that the average degree of preferential accretion in this case, $\dot{M}_2/\dot{M}_1$ over the final 500 orbits of the 2000-orbit simulation, is 7.06, compared to $\sim5.9$ for an $\alpha-$viscosity and $\sim4.4$ for a globally constant viscosity at the same Mach number and mass ratio. While this can not rule out that the viscosity profile within the cavity plays some additional role, it further demonstrates that the spatial viscosity profile governs accretion onto the binary. 

\subsubsection{The rate of accretion onto low-mass objects}
As shown in Figure \ref{fig:machsurvey}, we find that at $q\lesssim0.03$ the secondaries accrete more preferentially in colder disks.
The rate at which mass accretes through some surface can often be parameterized by $\dot{M}\sim\rho_c A_c v_c$, where $\rho_c,~A_c,$ and $v_c$ are characteristic densities, areas, and velocities respectively. When the flow of gas around an object is spherically symmetric, the characteristic speed is typically taken to be the sound speed $c_s$, and the area to be the surface area of a sphere with a radius given by $R_B=2GM/c_s^2$, roughly balancing thermal and gravitational energy, resulting in the Bondi accretion rate $\dot{M}_B\propto\rho M^2c_s^{-3}$ \citep{1952MNRAS.112..195B}. However, when the Hill radius of an embedded object becomes smaller than the Bondi radius, the surface area of the Hill sphere becomes a natural characteristic area. Similarly, when the Hill sphere becomes larger than the disk scale height, the characteristic area becomes $2\pi Hr_H$. The latter regime characterizes all of the binaries studied herein. 

Thus, we might expect an accretion rate onto the secondary along the lines of $\dot{M}\propto \rho H r_H v_c\propto\Sigma r_H v_c$. Whether the characteristic velocity is taken to simply be the sound speed, or related to the Hill radius and Keplerian shear as in \citet{2020MNRAS.498.2054R}, the accretion rate onto the secondary increasing as disks cool contradicts these expectations. We note that such predictions appear to be fairly accurate when measuring the rate at which material enters the sphere of influence of objects in disks (defined directionally, only measuring matter which is moving towards the secondary rather than that moving away from it) \citep{2023arXiv230501684C}. However, we have measured not just the rate at which gas enters the sphere of influence of the secondary, but the rate at which it accretes onto the secondary. Indeed, it appears that the accretion streams onto the secondary are more pronounced in higher Mach number disks, and that the corresponding cavities are larger and deeper, preventing accretion onto the primary, most likely because of weak shocks which become stronger in cooler disks. 

\subsection{Variability}\label{sec:variability}
\begin{figure}
\includegraphics[width=\linewidth]{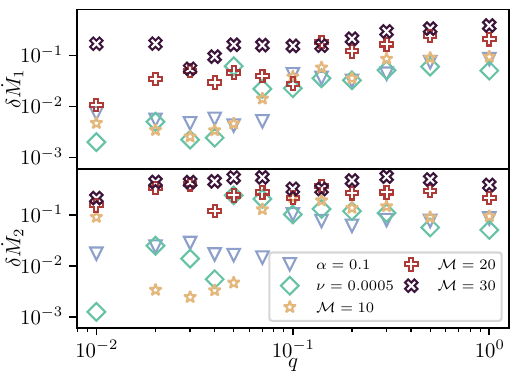}
\caption{The relative standard deviation of the accretion rate onto the primary ($\delta\dot{M}_1$) and onto the secondary ($\delta\dot{M}_2$). Accretion is never steady in an absolute sense. However, in cases where the cavity eccentricity is below $e\lesssim10^{-3}$ or so, the level of normalized variability is below $\sim1\%$. At most mass ratios, the degree of variability is greater for binaries in colder disks.}
\label{fig:variability}
\end{figure}
In the high mass ratio ($q\gtrsim0.2$) regime the accretion rate onto the binary is often governed by interactions with the pronouncedly eccentric cavities of their circumbinary disks and the overdense regions on the inner cavity edge, leading to modulation of the accretion rate on the orbital period of the binary (or half, in the case of equal-mass binaries) as binary members pass by the cavity pericenter, on the orbital period of the inner edge of the cavity and its overdensites, and the associated harmonics and beat frequencies \citep[e.g.,][]{2008ApJ...672...83M,2012ApJ...749..118S,2012ApJ...755...51N,2013MNRAS.436.2997D,2015ApJ...807..131S,2017MNRAS.466.1170M, 2019ApJ...879...76B,2020ApJ...889..114M,2021ApJ...921...71D,2021ApJ...922..175N,2022PhRvD.106j3010W,2022ApJ...928..187C,2023arXiv230402575M}.

Some studies \citep[e.g.,][]{2020ApJ...901...25D} have argued that below $q\lesssim0.2$ accretion variability on supraorbital timescales disappears, or that below $q\lesssim0.04$ the accretion is characterized by steady-state behavior rather than fluctuations \citep[e.g.,][]{2016MNRAS.459.2379D}: based on our higher-fidelity simulations,\footnote{Besides using cells near $r\sim a_b$ about $\sim12$ times smaller than \citet{2016MNRAS.459.2379D} and $\sim6$ times smaller than \citet{2020ApJ...901...25D}, and in addition to third- rather than second-order spatial reconstruction, we have also used more robust sink methods and time-averaged relevant quantities between outputs rather than outputting point values in time. Notably, \citet{2021ApJ...921...71D} illustrated that the sink methods used in prior studies led to spurious accretion variability.} we find that significant variability persists at least to $q=0.01$, and that in some cases this can occur on the orbital period of the cavity rather than that of the binary. To illustrate this, we show in Figure \ref{fig:timeseries} a sequence of accretion rate time series, at $q\in\{0.01, 0.03, 0.1, 0.3, 1.0\}$ and $\mathcal{M}\in\{10,20,30\}$. Evidently, below $q\lesssim0.1$, the accretion rate onto the primary, and its variability, become negligible compared to that onto the secondary. However, in most cases the accretion rate onto the secondary is still highly variable. Noting that at $\mathcal{M}=10$, the secondary of the $q=0.01$ binary exhibits accretion variability on the orbital period of the cavity, while the $q=0.03$ binary exhibits no appreciable variability, it appears, based on Figure \ref{fig:machsurveyDisk}, that the critical disk eccentricity below which the accretion flow becomes steady, if one exists, is $e \lesssim 10^{-3}$. Although the orbital period of the cavity dominates the accretion rate periodicity of the the $q=0.01$, $\mathcal{M}=10$ binary, oftentimes the most prominent period at $q\lesssim0.1$ is that of the binary orbit, although this is not to say that other periods are not present.

In an absolute sense, the accretion flow onto neither object is ever steady. However, in a number of low-mass ratio cases exhibit only very minute variability, to the point of being nearly imperceptible in visualizations such as Figure \ref{fig:timeseries}. To more quantitatively assess variability, we have calculated for each binary the relative standard deviation (also known as the coefficient of variation) for the accretion rate onto each object ($\delta M_i$), the standard deviation divided by the mean of each time series. Figure \ref{fig:variability} displays this metric for each accreting object. Thinner disks typically exhibit fluctuations on the order of $\sim10\%$ or more for at least the secondary, if not both objects. Thicker, $\mathcal{M}=10$ disks, on the other hand, can exhibit appreciably less accretion variability, particularly at $q\leq0.05$, although the $q=0.01$ binary accreting from a $\nu=0.001$ disk still exhibits accretion rate variability at the $\sim10\%$ level. While these low-variability cases often correspond to binaries with low-eccentricity disks ($e\lesssim10^{-2}$, Figure \ref{fig:viscsurveyDisk}), other disks - both thick and thin - with similarly small eccentricities lead to orders of magnitude greater variability at the same mass ratios.

\section{Discussion}\label{sec:discussion}
Our simulations provide insight into the evolution high-mass protoplanets, supermassive black hole binaries, and stellar binaries, among other systems. 
While the survey was performed at high resolution and covered a range of Mach number, disk viscosity, and mass ratio, many limitations remain. In particular we neglected binary eccentricity, used the overly simplistic locally isothermal equation of state, and, most importantly, performed the simulations using only two dimensional hydrodynamics.  That being said, these calculations can still provide general insight into a number of astrophysical systems and their dependence on physical parameters.
\subsection{Stars and Planets}
The Gaia mission has unveiled a substantial population of wide binaries stars with separations reaching $\sim1000~\rm{au}$ and mass ratios near unity ($q\gtrsim0.95$) \citep[e.g.,][]{2019MNRAS.489.5822E}, many of which are thought to have formed in and been shaped by circumbinary disks \citep{2022ApJ...933L..32H}. While accretion from all but the hottest, thickest disks \citep[such as those studied by][for example]{2005ApJ...623..922O} should cause the binary mass ratio to approach unity, the widening of binaries appears to have more sensitive dependencies. For thicker, $\mathcal{M}\lesssim10$ disks, it appears that, as discussed in Section \ref{sec:thickecc}, binaries that form with mass ratios $q\gtrsim0.1-0.3$ may spiral outward due to interactions with their circumbinary disk, depending on the effective disk viscosity. 

Concerning the interactions between planets and circumstellar disks, gravitational torques in the linear, very low mass regime have been thoroughly characterized, oftentimes leading to negative torques on the planet \citep[e.g.,][]{1979ApJ...233..857G,2002ApJ...565.1257T,2009MNRAS.394.2283P}. Further complications arise as planets become more massive, strongly perturbing the gas around them and potentially exciting eccentricity in the disk \citep[e.g.,][]{2006A&A...447..369K,2018ApJ...861..140K,2021ApJ...918L..36D}. In the fairly high-mass regime studied here, relevant to the evolution of brown dwarfs and super-Jupiters ($0.01\lesssim q \lesssim 0.1$), we have identified both positive torques, associated with the presence of long-lived precessing eccentric disks (correlating with but not depending on the amplitude of the disk eccentricity), and negative torques which transpire in the absence of these modes. The positive torques identified at $q\sim0.01$ may help planets formed through core accretion to move outwards to large separations, and outward migration may facilitate giant planet formation from otherwise under-massive cores \citep[e.g.,][]{2014ApJ...786...21P}.

\subsection{The Evolution and Detectability of SMBH binaries}
Studies of equal-mass binaries have suggested that thin accretion disks, as often invoked for active galactic nuclei as predicted by simple models of optically thick disks accreting at below the Eddington rate \citep[e.g.,][]{1973A&A....24..337S},\footnote{Notably, near-Eddington or very below-Eddington accretion rates are expected to beget fairly thick accretion disks \citep[e.g.,][]{1995ApJ...452..710N,2011A&A...527A..17S}. Disks with sufficient heating to avoid gravitational instability at large radii are also oftentimes predicted to be relatively thick \citep[e.g.,][]{2003MNRAS.341..501S,2020MNRAS.493.3732D,2022ApJ...928..191G}} have identified very strong negative torques, which could potentially drive binaries into the gravitaitonal wave-dominated regime of orbital evolution within the duration of a typical accretion episode \citep[][]{2022MNRAS.513.6158D}. However, as binary mass ratios decrease, these torques become appreciably weaker: at $\mathcal{M}=20$, $d\log{a}_b/d\log{M}$ increases from $\sim -1.5$ at $q=1$ to $1.2$ at $q=0.2$; and at $\mathcal{M}=30$, $d\log{a}_b/d\log{M}$ increases from $\sim-6.5$ at $q=1$ to $\sim-2.2$ at $q=0.1$. In some cases, for binaries around which eccentric disk modes are suppressed may inspiral at even faster rates, although we have not determined the reason behind this behavior.  While near-equal-mass binaries formed from major mergers may inspiral quickly and those with mass ratios $q\sim0.3$ more slowly, there is a chance that those with mass ratios $q\sim0.05$, like those predicted to result most frequently from galaxy mergers (e.g., \citet{2022arXiv220306016A,2019MNRAS.482.2913B}, but see also \citet{2018ApJ...864L..19T}), inspiral extremely quickly ($d\log{a}_b/d\log{M}\sim-35$). However, we also find that at lower mass ratios, $q\lesssim0.04$, binaries often outspiral due to interactions with circumbinary disks, which may cause such binaries to stall. 

Active galactic nuclei are often inferred to accrete at rates between $\sim1$ and $\sim1/100$ times the Eddington limit \citep[e.g.,][]{2008ApJ...680..169S}. Assuming that this holds true for binary systems as well, preferential accretion onto the secondary by the factors of $\sim2-7$ typically observed herein suggests that the more massive of the SMBHs typically accretes at an extremely sub-Eddington rate, due to both the tenuous supply of gas and higher mass. The resulting circumprimary accretion flows may be advection-dominated \citep[e.g.,][]{1995ApJ...452..710N}, or emit much more subtle radiation, making detection more challenging. We also note that because unequal-mass binaries often exhibit prominent accretion rate variability on their orbital period, such variations may be easier to robustly detect than the longer-timescale cavity-dominated variability that dominates in near-equal-mass binaries. 

\subsection{The importance of viscosity}
Interestingly, the functional form of the Navier-Stokes viscosity strongly affects the rate of preferential accretion onto the secondary (Figure \ref{fig:viscsurvey}), and at some mass ratios can affect the rate and direction of the evolution of the binary semi-major axis (Figure \ref{fig:viscsurveyLogs}). Although this clarifies some of the differences between prior investigations into the problem of unequal-mass binaries \citep[e.g.,][]{2020ApJ...889..114M,2021ApJ...921...71D}, it is somewhat unfortunate, because neither $\alpha-$viscosity nor a constant kinematic viscosity (of the magnitude considered herein) is truly present in astrophysical accretion disks: because the results of these simplified simulations are dependent on the viscous model employed, simulations with realistically driven angular momentum transport, via processes such as magnetohydrodynamic turbulence \citep[e.g.,][]{velikhov59} or gravitational instability \citep[e.g.,][]{2001ApJ...553..174G,2015ApJ...804...62R} will be necessary to assuredly understand preferential accretion, although in a broad sense the evolution of the binary semi-major axis appears less sensitive to viscous model assumptions. Recent advancements in numerical techniques may facilitate such studies, particularly for studies of the late inspirals of binary black holes \citep[e.g.,][]{2023arXiv230518538A}, and novel stochastic viscosity prescriptions may also hold promise \citep{2023arXiv230607199T}.

\subsection{Disk Eccentricity and Orbital Stability}
In a number of cases, examined in Section \ref{sec:nomode}, the circumbinary disk fails to develop a well-defined eccentric mode throughout the disk, an absence that is correlated strongly with changes in accretion morphology, leading to pronouncedly negative torques on the binary. Similar observations have been made previously, albeit with less precision \citep[e.g.,][]{2021ApJ...918L..36D}, the key difference being that disks may have smaller-magnitude eccentricities but still possess quasi-global eccentric modes and exhibit positive gravitational torques. Additionally, although disks with low eccentricities oftentimes coincide with small degrees of accretion variability, as shown in Section \ref{sec:variability}, this is by no means universal. 

Orbits about the triangular Lagrange points in the circular restricted three-body problem are linearly unstable for binary mass ratios $q>(25-\sqrt{621})/2\approx0.04$ \citep[e.g.,][]{1988fcm..book.....D}, and for this reason, we focused disproportionately on proximate mass ratios. Examining Figures \ref{fig:machsurvey} to \ref{fig:viscsurveyDisk}, and \ref{fig:variability}, the only changes we see associated specifically with $q=0.04$ occur in colder ($\mathcal{M}=20$, $\mathcal{M}=30$) disks, where eccentric modes are present at $q=0.03$ but not at $q=0.05$. 

A previous paper, \citet{2016MNRAS.459.2379D} 
suggested that the above instability threshold has drastic consequences, causing a transition from a steady flow at $q\lesssim0.04$ to a strongly fluctuating flow above that mass ratio. \citet{2016MNRAS.459.2379D} also suggested that this critical mass ratio marked a transition from a nearly-circular circumbinary disk with a narrow annular gap below $q\approx0.04$ to an eccentric disk with a large hollow cavity above. Our $\mathcal{M}=10$ simulations bear some resemblance to this picture, although they seem appear to change gradually below $q\lesssim0.1$. 
However, our simulations of thinner disks display qualitatively different behaviour, maintainting eccentric disks and accretion rate fluctuations greater than 10\% even down to $q=0.01$. However, we do confirm that in a broad sense, disk eccentricity, cavity size, and disk precession rate decrease along with the binary mass ratio. 

\section{Summary}\label{sec:summary}
We have carried out a broad survey of circumbinary accretion, studying circular binaries with mass ratios between $0.01$ and $1$; disk aspect ratios between $0.1$ and $0.0\bar{33}$ (azimuthal Mach numbers between $10$ and $30$), using both a constant-$\nu$ viscosity prescription (with $\nu=0.0005a_b^2\Omega_b$ and $\nu=0.001a_b^2\Omega_b$) and an $\alpha$ viscosity prescription. Although two-dimensional, our simulations have used very high resolutions, down to $\delta x \sim 0.00165a_b$, and third-order spatial reconstruction in addition to robust sink methodology - in fact, our tests in Appendix \ref{app:convergence} indicate that without using torque-free sinks \citep{2020ApJ...892L..29D,2021ApJ...921...71D}, it is \emph{impossible} to achieve converged results. Our main results are:
\begin{itemize}
     \item The spatial dependance of viscosity strongly determines the relative distribution of accretion between primary and secondary, likely because the more centrally concentrated surface density profiles reduces the amount of gas which flows from the Roche lobe of the secondary to that of the primary. 
     \item Cold ($\mathcal{M}\geq20$) disks that do not develop quasi-global precessing eccentric modes deliver very strong negative gravitational torques to the binary.
     \item Cavity eccentricity and the variability of accretion onto the binary broadly correlate with one another, and cavity eccentricities below $e\sim10^{-3}$ are often needed to preclude variability.
     \item The strong negative gravitational torques observed for circular binaries and thin disks become weaker as the binary mass ratio decreases.
     \item We find little evidence at these Mach numbers and viscosities that the stability of orbits about the fourth and fifth Lagrange points in the circular restricted three body problem plays a significant role in the dynamics of circumbinary accretion.
\end{itemize}
These results should further our understanding of the orbital evolution of giant planets, binary stars, and supermassive black holes.

\section*{Software}

\texttt{Athena++} \citep{2020ApJS..249....4S}, \texttt{matplotlib} \citep{4160265}, \texttt{cmocean} \citep{cmocean}, \texttt{numpy} \citep{5725236}, \texttt{yt} \citep{2011ApJS..192....9T}, \texttt{Disco} \citep{2016ApJS..226....2D,2021ApJ...921...71D}

\section*{Acknowledgments}
We wish to thank Adam Dempsey, Dan D'Orazio, Paul Duffell, and Cole Miller for discussions during the course of this project and feedback on a draft of this paper. 
AJD is also grateful for many useful discussions that took place at the Summer 2022 Center for Computational Astrophysics N-Body Workshop, especially with Jillian Bellovary.

The authors acknowledge the University of Maryland supercomputing resources (http://hpcc.umd.edu) that were made available for conducting the research reported in this paper, and the ASTRA cluster administered by the Center for Theory and Computation within the University of Maryland Department of Astronomy. Research at Perimeter Institute is supported in part by the Government of Canada through the Department of Innovation, Science and Economic Development and by the Province of Ontario through the Ministry of Colleges and Universities. AJD was supported in part by NASA ADAP grant 80NSSC21K0649. AJD gratefully acknowledges the support by LANL/LDRD under project number 20220087DR. The LA-UR number is LA-UR-23-27442. 

\appendix

\section{Convergence}\label{app:convergence}
We have performed a small set of convergence tests at mass ratios $q=0.01$ and $q=0.1$, the results of which suggest that our fiducial simulation parameters are sufficient to derive sufficiently converged results. First, at $q=0.1,~\mathcal{M}=10,~\nu=0.001\Omega_ba_b^2$ we performed and additional simulation at a base resolution of $512$ cells in both the $x$ and $y$ directions, resulting in a resolution near the binary of $\Delta x\approx0.0049a_b$. As a cross-check between our different refinement schemes, we also performed a simulation with the aforementioned physical parameters but the grid, sink, and softening configuration employed in our lower mass ratio simulations. The results of these tests are displayed in Table \ref{tab:convergence}. The total torque on the binary in very well converged, varying by $<1\%$ between simulations. At constant gravitational softening length, the degree of preferential accretion is also very well converged with resolution, varying by $<1\%$. However, $\dot{M}_2/\dot{M}_1$ appears to depend fairly strongly on the gravitational softening length, decreasing as the softening length scale decreases. One possible explanation for this trend is that, similarly to $\alpha-$viscosity but in this case because of the decreased shear in the inner disk, gravitational softening leads to a steady-state surface density profile which decreases with radius which may reduce the rate of mass transfer between the Roche lobe of the primary and secondary. 

\begin{figure}
\includegraphics[width=\linewidth]{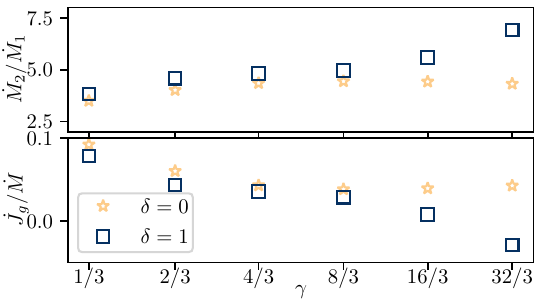}
\caption{The effect of sink rate on the degree of preferential accretion. At sink rates that are too slow, mass builds up regardless of the sink method used, leading to erroneous torques and preferential accretion measurements. Once sufficiently fast sinks are employed, torque-free ($\delta=0$) sinks converge, while standard ($\delta=1$) sinks, if sufficiently fast, can lead to spurious results of any magnitude. }
\label{fig:rates}
\end{figure}

\begin{deluxetable}{cccccc}
\tablehead{ \colhead{$q$} & \colhead{$\mathcal{M}$} & \colhead{Resolution} & \colhead{$\epsilon_{g,2}/a_b$} & \colhead{$\dot{M}_2/\dot{M}_1$} & \colhead{$\dot{J}/\dot{M}$}}
\startdata
0.1 & 10 & 384 & 0.035 & 4.44 & 0.714 \\
0.1 & 10 & 512 & 0.035 & 4.46 & 0.719 \\
0.1 & 10 & 512 & 0.025  & 4.15 & 0.720 \\
0.1 & 10 & $\epsilon=4/300$ & 0.019  & 3.51 & 0.712 \\
0.01 & 10 & $\epsilon=4/300$ & 0.0094 & 2.84 & 0.806 \\
0.01 & 10 & $\epsilon=1/100$ & 0.0094 & 2.88 & 0.809 \\
\hline
\enddata
\caption{The results of our resolution tests. The torque on the binary is converged regardless of numerical parameters. However, the degree of preferential accretion depends fairly strongly on the gravitational softening used in each simulation.  }\label{tab:convergence}
\end{deluxetable}
Because certain mass removal algorithms and parameter choices can lead to unphysical biases \citep{2020ApJ...892L..29D,2021ApJ...921...71D}, we test the extent to which our choices of sink rate $\gamma_i$ affect our results. Although our primary suite of simulations all used torque free sinks, setting $\delta=0$ in Equation (\ref{eq:tf2}), we carried out a limited number of $\nu=0.001,~q=0.1,~\mathcal{M}=10$ simulations to test the extent to which use of standard ($\delta=1$) sinks may have contributed to discrepancies in the values of $\dot{M}_2/\dot{M}_1$ previously reported in the literature \citep[e.g.,][]{2014ApJ...783..134F,2020ApJ...889..114M,2022arXiv220302514S}. Time-averaged values of $\dot{M}_2/\dot{M}_1$ and $\dot{J}_g/\dot{M}$ are shown in Figure \ref{fig:rates}. Although the values measured in simulations which used torque-free sinks are broadly consistent with one another, results derived using standard sinks depend strongly on the sink rate. At low sink rates, torque-free and standard sinks are in better agreement, but if the sink rate is too low (in this case $\gamma\lesssim4/3$) the results become erroneous, as the sink is unable to remove mass sufficiently quickly and artificial density gradients develop.

\begin{figure}
\includegraphics[width=\linewidth]{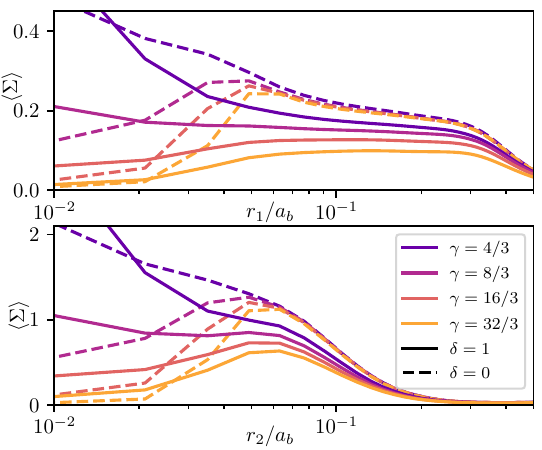}
\caption{Time- and azimuth-averaged surface density profiles in the frame of the primary (top panel) and secondary (bottom panel). Different line colors, from dark to light, indicate different sink rates from slow to fast. Solid lines plot profiles derived from simulations using standard sinks and dashed lines plot profiles derived from simulations using torque-free sinks, which are the default for the simulations in this work. We observe that standard sinks significantly deplete the surface density profile well beyond the sink radius ($0.035a_b$), and the profiles do not converge at any value of the sink rate. Torque-free sinks, on the other hand, appear to produce converged surface density profiles throughout the bulk of the minidisks at $\gamma\geq8/3$.}
\label{fig:miniprofs}
\end{figure}

Figure \ref{fig:miniprofs} illustrates time-averaged density profiles of the minidisks in these simulations. Notably, at slower sink rates, the surface density profile from the standard-sink simulations approaches that of the simulations which used torque-free sinks. Although the simulations using standard sinks display no semblance of convergence with sink rate, the surface density profiles in simulations using torque-free sinks and sink rates $\gamma\geq8/3$ approximately converge far from the sinks. Furthermore, the surface density profile around the primary in the $\gamma=4/3$ simulation, which deviates from the others well beyond the sink region, suggests that mass is being removed very slightly too slowly, allowing the disk structure outside of the sink to be affected. Thus, we consider both our default algorithmic choices and sink rates to be well-justified. 

We performed a similar test at $q=0.01$, for a disk with $\nu=0.001$ and $\mathcal{M}=10$. Using out fiducial resolution, sink size, and softening parameters, we tested the sink rates $\gamma_0=\{4/3, 8/3, 16/3\}$. Taking our fiducial sink rate, $\gamma_0=8/3$ as a baseline, we found that the accretion rate ratio $\dot{M}_2/\dot{M}_1$ and gravitational torque $\dot{J}_g/\dot{M}$ agreed to better than $3\%$ between the $\gamma_0=1.33$ and $\gamma_0=2.67$ runs, and agreed to better than $1\%$ between the $\gamma_0=2.67$ and $\gamma_0=5.33$ runs, suggesting that our fiducial choice of $\gamma_0=1.33$ is appropriate. 

\section{Comparisons with previous work}\label{app:compare}
Over the last few years, a few studies have investigated the orbital evolution of binaries with $q\sim10^{-2}$ and found results in disagreement with what we have reported in this work. Namely, when nominally studying binaries with $10^{-2}\leq q \leq 1$, in $\mathcal{M}=10$, $\nu=0.001$ disks, \citet{2020ApJ...901...25D} found not only that binaries inspiral at $q<0.1$ - where we find that binaries should outspiral - but also values of $\dot{M}_2/\dot{M}_1$ appreciably higher than in our simulations. Additionally, \citet{2021ApJ...918L..36D} reported negative gravitational torques in a simulation of a $q=0.01$ binary with a $\mathcal{M}=10$ disk. Below, we identify some of the modeling choices responsible for these discrepancies. 

\subsection{Comparison with Duffell et al. 2020}\label{sec:paulcomp}
The present study differs from \citet{2020ApJ...901...25D} in a number of ways: the code utilized (\texttt{Athena++} vs \texttt{Disco}), the simulation initial conditions, the simulation diagnostics, and the fact that \citet{2020ApJ...901...25D} solved a fundamentally different set of equations describing viscous hydrodynamics. To investigate this, we have carried out a small suite of simulations using different versions of the \texttt{Disco} code. We find results which are consisten with our main, \texttt{Athena++}-derived, results when solving the same set of equations and reporting similar diagnostic quantities. 

The most important factor contributing to the differences in the values of $\dot{M}_2/\dot{M}_1$ measured in our simulations and those in \citet{2020ApJ...901...25D} is that we solved the Navier-Stokes (NS) equations, given by Equation (\ref{eq:momentum}), with the velocity shear tensor defined by
\begin{equation}\label{eq:velshear}
\sigma_{ij} = \frac{1}{2}(\nabla_iv_j + \nabla_jv_i)+\frac{1}{d}g_{ij}\nabla_kv^k,
\end{equation}
where $g_{ij}$ is the metric tensor, superscripts ($^i$) denote contravariant tensor components, subscripts ($_i$) denote covariant tensor components, and $d=3$. However, the version of \texttt{Disco} used in \citet{2020ApJ...901...25D} solved a modified set of equations, where Equations (\ref{eq:momentum}) and (\ref{eq:velshear}) are simplified under the assumption of a globally constant dynamic viscosity ($\nu\Sigma$) and $d=2$, where the former assumption is strongly violated in typical circumbinary disks (see, e.g. Figure \ref{fig:gridBig}). A derivation of this set of modified Navier-Stokes (NS$^*$) equations, and a comparison with the standard NS equations, was presented in Appendix A of \citet{2021ApJ...921...71D}. We denote an earlier version of \texttt{Disco},\footnote{\url{https://github.com/duffell/Disco}. In this work we have used commit 833bb65ab937bc8e18d109cd87afa95ddda5880c.} which only solves the NS$^*$ equations, as \texttt{Disco v1}. As shown in Table \ref{tab:paulcomp}, simulations solving the NS$^*$ equations find systematically higher values of $\dot{M}_2/\dot{M}_1$, which also depend more strongly on numerical dissipation than results from simulations solving the NS equations.

Some more recent studies \citep{2021ApJ...921...71D,2022MNRAS.513.6158D} have used an updated version of \texttt{Disco}, which solves the NS equations (as well as the NS$^*$ equations, for the sole purpose of making comparisons to \texttt{Disco v1}), and includes a number of other upgrades such as using cell centroids rather than centres where appropriate and time-averaging outputs every time step properly throughout the Runge-Kutta time integration. We denote this version of \texttt{Disco} simply as \texttt{Disco}.\footnote{We used \url{https://github.com/ajdittmann/Disco/}, commit 9225403b01a4c5634e9a21e05615d91e72de3e99, but see also \url{https://github.com/NYU-CAL/Disco} for a more up-to-date version.}

\setlength{\tabcolsep}{2.2pt}
\begin{deluxetable}{ccccccccc}
\tablehead{
\colhead{Code} & \colhead{Eqs.} & \colhead{q} &\colhead{$N_r$} & \colhead{$\theta_{\rm PLM}$} & \colhead{$\langle\dot{M}_2\rangle/\langle\dot{M}_1\rangle$} & \colhead{$\langle\dot{J}_{g,1}\rangle$} & \colhead{$\langle\dot{J}_{g,2}\rangle$} & \colhead{$\langle\dot{J}_g\rangle$}}
\startdata
$\texttt{Disco}$&NS&0.02&512&1&3.303 & -0.021 & 0.096 & 0.075 \\
$\texttt{Disco}$&NS&0.02&512&1.5&2.951 & -0.021 & 0.133 & 0.112 \\
$\texttt{Disco}$&NS&0.02&768&1.5&3.126 & -0.023 & 0.132 & 0.108 \\
$\texttt{Disco}$&NS$^*$&0.02&512&1&6.145 & -0.030 & 0.042 & 0.011 \\
$\texttt{Disco}$&NS$^*$&0.02&512&1.5&6.396 & -0.033 & 0.027 & -0.006 \\
$\texttt{Disco}$&NS$^*$&0.02&768&1.5&9.870 & -0.034 & -0.013 & -0.048 \\
$\texttt{Disco~v1}^*$&NS$^*$&0.02&512&1&5.235 & -0.027 & 0.058 & 0.031 \\
$\texttt{Disco~v1}^*$&NS$^*$&0.02&512&1.5&6.691 & -0.030 & 0.020 & -0.010 \\
$\texttt{Disco~v1}^*$&NS$^*$&0.02&768&1.5&8.061 & -0.032 & 0.005 & -0.027 \\
$\texttt{Disco~v1}$&NS$^*$&0.02&512&1&5.235 & 0.058 & -1.367 & -1.309 \\
$\texttt{Disco~v1}$&NS$^*$&0.02&512&1.5&6.691 & 0.020 & -1.525 & -1.505 \\
$\texttt{Disco~v1}$&NS$^*$&0.02&768&1.5&8.061 & 0.005 & -1.579 & -1.574 \\
\enddata
\label{tab:paulcomp}
\caption{The results from a series of simulations of accretion onto $q=0.02$ binaries from $\mathcal{M}=10$, $\nu=0.001$ disks, using different versions of \texttt{Disco}, solving both the Navier-Stokes equations and the set of equations solved in \citet{2020ApJ...901...25D}, and at different levels of numerical dissipation. Gravitational torques are presented in units of $3\pi\nu\Sigma_0a_b^4\Omega_b$, as in \citet{2020ApJ...901...25D}, rather than units of $\dot{M}a_b^2\Omega_b$ as in every other section of this paper. For our simulations using \texttt{Disco}, we report per-timestep time averages of each quantity over the last 500 binary orbits; because \texttt{Disco v1} lacks this functionality, for those simulation we report the mean of point values which were output 100 times per orbit, over the last 500 orbits. When using \texttt{Disco} to solve the Navier-Stokes equations, we find consistent results with our \texttt{Athena++} simulations. When solving a modified equation set and reproducing errors in torque calculations, we are able to reproduce the results of \citet{2020ApJ...901...25D}.}
\end{deluxetable}

Our simulations using both \texttt{Disco} and \texttt{Disco v1} indicate that the largest differences between the gravitational torques measured in our \texttt{Athena++} simulations and those reported in \citet{2020ApJ...901...25D} are due to errors in the simulation diagnostics used in \citet{2020ApJ...901...25D}, which we have identified in the \texttt{Disco v1} source code.\footnote{See \url{https://github.com/duffell/Disco/blob/833bb65ab937bc8e18d109cd87afa95ddda5880c/report.c\#L149} and \url{https://github.com/duffell/Disco/blob/833bb65ab937bc8e18d109cd87afa95ddda5880c/Planets/bin_varyq.c}. We note that \texttt{Disco v1} names the torque on the secondary ``Torque,'' and the torque on the primary ``Torque2.''}
Specifically, \texttt{Disco v1} uses the lever arm of the secondary, in the barycentric frame, to calculate the torque on both the secondary and the primary. 
Our code comparison and source code examination also suggests that \citet{2020ApJ...901...25D} mislabeled the gravitiatonal torques on the primary and secondary, matching the variable names in the source code.
We denote a verison of \texttt{Disco v1} that corrects these typos as \texttt{Disco v1}$^*$. 
As shown in Table \ref{tab:paulcomp}, we find results in good agreement with \citet{2020ApJ...901...25D} and the base version of \texttt{Disco v1}. However, if we correct the typos in the torque calculation and reporting, we find good agreement between \texttt{Disco v1}$^*$, \texttt{Athena++}, and \texttt{Disco}.

When attempting to compare our results to those of \citet{2020ApJ...901...25D}, we have only focused on a single mass ratio, $q=0.02$ rather than adiabatically varying the mass ratio. We use the same initial conditions described in Section \ref{sec:setup}, albeit using a different mesh: rather than a Cartesian domain extending from $-10a_b$ to $10a_b$ in $x$ and $y$ with static mesh refinement as in our \texttt{Athena++} simulations, we use a cylindrical domain which extends from $r=0$ to $r=30$, with approximately uniform resolution within $r<a_b$ and approximately log-uniformly spaced cells from $a_b\leq r\leq30a_b$, setting the number of azimuthal cells at each radius so that $r\Delta\phi\approx\Delta r$. All simulations presented here used the HLLC Riemann solver and set the motion of each grid annulus to match the average azimuthal velocity of the fluid elements therein. To test the effects of numerical dissipation, particularly in the NS$^*$ formalism, we tested multiple resolutions ($N_r=512$, as in \citet{2020ApJ...901...25D}, and $N_r=768$) as well as different slope limiting coefficients $\theta_{\rm PLM}$, where $\theta_{\rm PLM}=1$ corresponds to a more numerically dissipative approximation of the slope in a piecewise linear reconstruction and $\theta_{\rm PLM}=1.5$ leads to a slightly less dissipative, but still total variation diminishing, reconstruction using a generalized minmod formulation \citep[see][]{2000JCoPh.160..241K,2016ApJS..226....2D}. For the sake of comparison, all of these simulations used standard ($\delta=1$) sinks with $\gamma=3$ and $r_s=0.05a_b$, as in \citet{2020ApJ...901...25D}. Thus, these results \textit{can not} be converged, but may at best appear so by chance (see Appendix \ref{app:convergence}). 

\subsection{Comparison with Dempsey et al. 2021}

\setlength{\tabcolsep}{11.2pt}
\begin{deluxetable}{ccc}
\tablehead{
\colhead{Accretion onto secondary} & \colhead{$\epsilon_{g,2}/a_b$} & \colhead{$\langle\dot{J}_g\rangle/\langle\dot{M}\rangle$} }
\startdata
Yes & 0.0095 & 0.0780 \\
No & 0.0095 & 0.1543 \\
Yes & 0.06 & 0.022 \\
No & 0.06 & -0.0782 \\
\enddata
\label{tab:adamcomp}
\caption{The results from a series of simulations attempting to reconcile the differences between our results and those of \citet{2021ApJ...918L..36D} for $q=0.01$ binaries. Both a large softening length \textit{and} disallowance of accretion onto the secondary are necessary to reproduce the results of \citet{2021ApJ...918L..36D}.}
\end{deluxetable}

One of the primary results of \citet{2021ApJ...918L..36D} was the transition from inspiral at lower mass ratios to outspirals at high mass ratios. More precisely, \citet{2021ApJ...918L..36D} found that, in terms of the gap-width scaling parameter $K'=q^2\alpha^{-1}h^{-1}$ \citep[the ratio of the one-sided gravitational torque in the absence of a gap to the viscous torque, scaled by the square of the disk aspect ratio $h=H/r$, e.g.,][]{1986ApJ...309..846L,2021ApJ...918L..36D}, binaries with $K'\lesssim10$ experience negative gravitational torques whereas those with $K'\gtrsim10$ experience positive gravitational torques (See Figure 2 in \citet{2021ApJ...918L..36D}). However, as an example, we have found positive torques in all of our simulations at $q=0.01,$ which have $K'\leq1$.

One factor which may have contributed to \citet{2021ApJ...918L..36D} finding negative gravitational torques in \textit{all} simulations with $K'\lesssim10$ is their omission, as detailed in their first Appendix, of terms in the gravitational torque related to performing simulations in the stellocentric frame. The neglected terms can be sufficiently large to lead to positive torques in some simulations for which they reported negative torques. 

Two other factors contributing to the difference in results between our study and \citet{2021ApJ...918L..36D} are that while we allowed both point masses to accrete and used gravitational softening values that were much smaller than the Hill radius of a given point mass, \citet{2021ApJ...918L..36D} did not allow accretion onto the secondary and also used much larger values of the softening, $0.06$ for $\mathcal{M}=10$ disks, following \citet{2012A&A...541A.123M}. We have conducted an additional set of simulations of $q=0.01,~\mathcal{M}=10,~\nu=0.001$ systems, the results of which are presented in Table \ref{tab:adamcomp}, and determined that \emph{both} larger gravitational softening values and disallowance of accretion onto the secondary are necessary to produce negative gravitational torques in such systems.

\section{Simulation Summaries}\label{app:summary}
For general convenience, we have collected a number of results concerning disk morphology and binary evolution from our main suite of simulations in Table \ref{tab:summary2}.

\renewcommand{\arraystretch}{0.825}
\setlength{\tabcolsep}{2.6pt}
\begin{deluxetable*}{cccccccccccccc}
\tablehead{
\colhead{viscosity} & \colhead{$\mathcal{M}$} & \colhead{$q$} & \colhead{$\dot{M}_2/\dot{M}_1$} & \colhead{$\dot{M}/\dot{M}_0 $}& \colhead{$\dot{J}_g/\dot{M}$} & \colhead{$\dot{J}_a/\dot{M} $} & \colhead{$\dot{J}/\dot{M}$} & \colhead{$d\log{a}_b/d\log{M}$} & \colhead{$dq/d\log{M}$} & \colhead{$a_{\rm cav}$} & \colhead{$e_{\rm cav}$} & \colhead{$r_p$} &\colhead{$\tau_\omega$} }
\startdata
$\alpha=0.1$&10&0.01&2.885 & 1.018 & 0.032 & 0.730 & 0.762 & 5.869 & 0.747 & 1.65 & 2.863e-05 & 1.650 & 1648.504\\
$\alpha=0.1$&10&0.02&3.990 & 1.042 & 0.043 & 0.768 & 0.811 & 3.391 & 0.811 & 1.75 & 8.837e-03 & 1.735 & -1071.889\\
$\alpha=0.1$&10&0.03&4.246 & 1.059 & 0.082 & 0.761 & 0.843 & 4.635 & 0.828 & 1.75 & 2.094e-05 & 1.750 & -1290.371\\
$\alpha=0.1$&10&0.04&4.806 & 1.049 & 0.059 & 0.765 & 0.824 & 2.166 & 0.854 & 1.75 & 2.738e-05 & 1.750 & -3201.226\\
$\alpha=0.1$&10&0.05&5.140 & 1.032 & 0.032 & 0.760 & 0.792 & 0.428 & 0.870 & 1.95 & 9.771e-05 & 1.950 & 9016.692\\
$\alpha=0.1$&10&0.07&5.406 & 0.993 & -0.028 & 0.738 & 0.710 & -1.901 & 0.891 & 1.95 & 1.707e-03 & 1.947 & 772.109\\
$\alpha=0.1$&10&0.1&5.906 & 1.019 & 0.048 & 0.709 & 0.758 & 0.201 & 0.925 & 2.05 & 1.583e-01 & 1.725 & 515.883\\
$\alpha=0.1$&10&0.14&6.562 & 1.017 & 0.073 & 0.670 & 0.743 & 0.365 & 0.968 & 2.15 & 1.594e-01 & 1.807 & 447.922\\
$\alpha=0.1$&10&0.2&6.772 & 0.981 & 0.038 & 0.608 & 0.647 & -0.454 & 1.015 & 2.15 & 1.482e-01 & 1.831 & 263.607\\
$\alpha=0.1$&10&0.3&4.618 & 1.025 & 0.221 & 0.497 & 0.718 & 1.499 & 0.999 & 2.65 & 1.974e-01 & 2.127 & 323.543\\
$\alpha=0.1$&10&0.5&2.279 & 1.019 & 0.335 & 0.343 & 0.679 & 2.022 & 0.814 & 2.85 & 2.069e-01 & 2.260 & 295.727\\
$\alpha=0.1$&10&1.0&0.999 & 1.018 & 0.415 & 0.250 & 0.665 & 2.322 & -0.001 & 2.85 & 2.113e-01 & 2.248 & 278.994\\
$\nu=0.0005$&10&0.01&2.715 & 1.070 & 0.087 & 0.716 & 0.804 & 16.808 & 0.735 & 1.65 & 6.588e-06 & 1.650 & -1982.662\\
$\nu=0.0005$&10&0.02&2.668 & 1.055 & 0.073 & 0.699 & 0.772 & 6.576 & 0.736 & 1.75 & 1.394e-02 & 1.726 & -859.748\\
$\nu=0.0005$&10&0.03&2.358 & 1.036 & 0.070 & 0.662 & 0.732 & 3.963 & 0.714 & 1.85 & 5.285e-04 & 1.849 & 0.000\\
$\nu=0.0005$&10&0.04&2.080 & 1.022 & 0.078 & 0.625 & 0.702 & 3.192 & 0.689 & 1.95 & 1.178e-03 & 1.948 & 734.292\\
$\nu=0.0005$&10&0.05&2.921 & 1.025 & 0.001 & 0.676 & 0.677 & -0.959 & 0.769 & 2.05 & 8.498e-02 & 1.876 & 755.319\\
$\nu=0.0005$&10&0.07&3.131 & 1.023 & 0.050 & 0.663 & 0.713 & 0.627 & 0.793 & 2.45 & 1.351e-01 & 2.119 & 746.407\\
$\nu=0.0005$&10&0.1&3.992 & 0.959 & -0.055 & 0.663 & 0.608 & -2.317 & 0.858 & 2.45 & 1.419e-01 & 2.102 & 594.494\\
$\nu=0.0005$&10&0.14&3.076 & 1.016 & 0.148 & 0.585 & 0.733 & 1.754 & 0.821 & 2.55 & 1.911e-01 & 2.063 & 569.975\\
$\nu=0.0005$&10&0.2&2.378 & 0.999 & 0.186 & 0.497 & 0.683 & 1.677 & 0.774 & 2.75 & 2.171e-01 & 2.153 & 565.187\\
$\nu=0.0005$&10&0.3&1.850 & 0.981 & 0.235 & 0.403 & 0.638 & 1.652 & 0.707 & 2.85 & 2.529e-01 & 2.129 & 613.830\\
$\nu=0.0005$&10&0.5&1.797 & 0.986 & 0.302 & 0.325 & 0.627 & 1.720 & 0.696 & 3.25 & 2.410e-01 & 2.467 & 544.394\\
$\nu=0.0005$&10&1.0&1.001 & 0.980 & 0.354 & 0.250 & 0.604 & 1.834 & 0.001 & 3.35 & 2.497e-01 & 2.514 & 548.888\\
$\nu=0.001$&10&0.01&2.842 & 1.049 & 0.078 & 0.729 & 0.806 & 15.526 & 0.744 & 1.65 & 1.000e-03 & 1.648 & 0.000\\
$\nu=0.001$&10&0.02&2.685 & 1.056 & 0.113 & 0.699 & 0.811 & 10.536 & 0.738 & 1.75 & 2.191e-05 & 1.750 & -1505.817\\
$\nu=0.001$&10&0.03&2.608 & 1.069 & 0.148 & 0.680 & 0.829 & 9.399 & 0.736 & 1.85 & 3.660e-05 & 1.850 & -2997.582\\
$\nu=0.001$&10&0.04&2.551 & 1.052 & 0.135 & 0.664 & 0.798 & 6.237 & 0.735 & 1.95 & 9.111e-05 & 1.950 & 7914.738\\
$\nu=0.001$&10&0.05&2.457 & 1.029 & 0.112 & 0.645 & 0.757 & 3.928 & 0.731 & 1.95 & 3.449e-04 & 1.949 & 1584.398\\
$\nu=0.001$&10&0.07&2.290 & 0.986 & 0.058 & 0.609 & 0.667 & 0.888 & 0.722 & 2.05 & 2.323e-02 & 2.002 & 604.878\\
$\nu=0.001$&10&0.1&4.440 & 1.018 & 0.038 & 0.676 & 0.714 & -0.077 & 0.878 & 2.45 & 1.382e-01 & 2.111 & 587.053\\
$\nu=0.001$&10&0.14&3.914 & 0.970 & -0.008 & 0.616 & 0.608 & -1.147 & 0.876 & 2.15 & 1.522e-01 & 1.823 & 412.126\\
$\nu=0.001$&10&0.2&3.841 & 0.913 & -0.096 & 0.557 & 0.461 & -2.376 & 0.903 & 2.15 & 1.336e-01 & 1.863 & 341.719\\
$\nu=0.001$&10&0.3&2.584 & 1.027 & 0.279 & 0.442 & 0.721 & 2.149 & 0.829 & 2.85 & 2.108e-01 & 2.249 & 414.573\\
$\nu=0.001$&10&0.5&1.706 & 1.025 & 0.380 & 0.321 & 0.702 & 2.425 & 0.669 & 2.75 & 2.484e-01 & 2.067 & 390.756\\
$\nu=0.001$&10&1.0&1.001 & 1.024 & 0.445 & 0.250 & 0.695 & 2.562 & 0.001 & 2.85 & 2.437e-01 & 2.156 & 372.150\\
$\nu=0.001$&20&0.01&6.073 & 1.406 & 0.058 & 0.830 & 0.888 & 8.505 & 0.866 & 1.63 & 8.973e-04 & 1.632 & 0.000\\
$\nu=0.001$&20&0.02&7.084 & 1.390 & 0.021 & 0.836 & 0.857 & 0.546 & 0.891 & 1.85 & 1.311e-01 & 1.607 & 2067.261\\
$\nu=0.001$&20&0.03&7.346 & 1.477 & 0.023 & 0.825 & 0.848 & 0.269 & 0.903 & 1.75 & 1.323e-01 & 1.521 & 1003.759\\
$\nu=0.001$&20&0.04&4.633 & 0.970 & -0.638 & 0.759 & 0.121 & -35.596 & 0.848 & 2.04 & 4.545e-04 & 2.039 & 0.000\\
$\nu=0.001$&20&0.05&3.820 & 0.976 & -0.584 & 0.717 & 0.133 & -26.870 & 0.821 & 1.95 & 1.234e-02 & 1.926 & 0.000\\
$\nu=0.001$&20&0.07&2.901 & 0.885 & -0.752 & 0.650 & -0.102 & -25.614 & 0.777 & 2.15 & 1.989e-02 & 2.107 & 0.000\\
$\nu=0.001$&20&0.1&2.566 & 0.909 & -0.709 & 0.597 & -0.112 & -18.163 & 0.761 & 1.95 & 2.475e-02 & 1.902 & 0.000\\
$\nu=0.001$&20&0.14&2.487 & 0.819 & -1.051 & 0.553 & -0.497 & -20.502 & 0.767 & 1.88 & 1.541e-02 & 1.847 & 0.000\\
$\nu=0.001$&20&0.2&4.991 & 1.290 & 0.151 & 0.583 & 0.734 & 1.176 & 0.960 & 2.85 & 3.142e-01 & 1.955 & 386.267\\
$\nu=0.001$&20&0.3&3.108 & 1.189 & 0.085 & 0.461 & 0.545 & -0.044 & 0.889 & 2.75 & 2.992e-01 & 1.927 & 280.629\\
$\nu=0.001$&20&0.5&1.959 & 1.104 & -0.013 & 0.332 & 0.319 & -1.113 & 0.740 & 2.75 & 2.829e-01 & 1.972 & 247.952\\
$\nu=0.001$&20&1.0&0.999 & 1.061 & -0.063 & 0.250 & 0.187 & -1.505 & -0.001 & 2.75 & 2.749e-01 & 1.994 & 229.242\\
$\nu=0.001$&30&0.01&6.972 & 1.696 & 0.033 & 0.846 & 0.879 & 3.431 & 0.882 & 1.75 & 7.179e-02 & 1.624 & 773.336\\
$\nu=0.001$&30&0.02&7.766 & 1.894 & 0.086 & 0.847 & 0.934 & 7.536 & 0.901 & 1.95 & 1.869e-01 & 1.586 & 976.336\\
$\nu=0.001$&30&0.03&6.351 & 1.795 & 0.143 & 0.814 & 0.958 & 9.117 & 0.886 & 2.16 & 2.465e-01 & 1.631 & 1603.165\\
$\nu=0.001$&30&0.04&5.911 & 1.809 & 0.169 & 0.794 & 0.963 & 8.323 & 0.884 & 2.43 & 2.176e-01 & 1.898 & 790.916\\
$\nu=0.001$&30&0.05&5.648 & 1.399 & -0.294 & 0.769 & 0.475 & -14.051 & 0.884 & 1.85 & 3.237e-02 & 1.790 & 0.000\\
$\nu=0.001$&30&0.07&4.232 & 1.307 & -0.368 & 0.706 & 0.338 & -13.094 & 0.851 & 1.85 & 2.342e-02 & 1.807 & 0.000\\
$\nu=0.001$&30&0.1&8.384 & 1.533 & -0.050 & 0.740 & 0.690 & -2.191 & 0.971 & 2.55 & 3.150e-01 & 1.747 & 341.975\\
$\nu=0.001$&30&0.14&5.907 & 1.384 & -0.209 & 0.661 & 0.452 & -4.866 & 0.952 & 2.55 & 3.070e-01 & 1.767 & 248.328\\
$\nu=0.001$&30&0.2&4.368 & 1.216 & -0.441 & 0.571 & 0.130 & -7.343 & 0.932 & 2.65 & 2.813e-01 & 1.904 & 208.370\\
$\nu=0.001$&30&0.3&2.986 & 1.089 & -0.725 & 0.457 & -0.268 & -9.169 & 0.876 & 2.65 & 2.760e-01 & 1.919 & 213.041\\
$\nu=0.001$&30&0.5&1.503 & 1.022 & -0.796 & 0.311 & -0.485 & -8.166 & 0.601 & 2.75 & 3.027e-01 & 1.918 & 246.446\\
$\nu=0.001$&30&1.0&0.993 & 1.041 & -0.682 & 0.250 & -0.432 & -6.455 & -0.007 & 2.85 & 3.051e-01 & 1.980 & 262.483\\
\enddata
\label{tab:summary2}\caption{The results of our main parameter survey}
\end{deluxetable*}

\bibliographystyle{aasjournal}
\bibliography{references}

\begin{thebibliography}{}
\expandafter\ifx\csname natexlab\endcsname\relax\def\natexlab#1{#1}\fi
\providecommand{\url}[1]{\href{#1}{#1}}

\bibitem[{{Agazie} {et~al.}(2023){Agazie}, {Anumarlapudi}, {Archibald},
  {Arzoumanian}, {Baker}, {Becsy}, {Blecha}, {Brazier}, {Brook},
  {Burke-Spolaor}, {Burnette}, {Case}, {Charisi}, {Chatterjee},
  {Chatziioannou}, {Cheeseboro}, {Chen}, {Cohen}, {Cordes}, {Cornish},
  {Crawford}, {Cromartie}, {Crowter}, {Cutler}, {DeCesar}, {DeGan}, {Demorest},
  {Deng}, {Dolch}, {Drachler}, {Ellis}, {Ferrara}, {Fiore}, {Fonseca},
  {Freedman}, {Garver-Daniels}, {Gentile}, {Gersbach}, {Glaser}, {Good},
  {Gultekin}, {Hazboun}, {Hourihane}, {Islo}, {Jennings}, {Johnson}, {Jones},
  {Kaiser}, {Kaplan}, {Kelley}, {Kerr}, {Key}, {Klein}, {Laal}, {Lam}, {Lamb},
  {Lazio}, {Lewandowska}, {Littenberg}, {Liu}, {Lommen}, {Lorimer}, {Luo},
  {Lynch}, {Ma}, {Madison}, {Mattson}, {McEwen}, {McKee}, {McLaughlin},
  {McMann}, {Meyers}, {Meyers}, {Mingarelli}, {Mitridate}, {Natarajan}, {Ng},
  {Nice}, {Ocker}, {Olum}, {Pennucci}, {Perera}, {Petrov}, {Pol}, {Radovan},
  {Ransom}, {Ray}, {Romano}, {Sardesai}, {Schmiedekamp}, {Schmiedekamp},
  {Schmitz}, {Schult}, {Shapiro-Albert}, {Siemens}, {Simon}, {Siwek}, {Stairs},
  {Stinebring}, {Stovall}, {Sun}, {Susobhanan}, {Swiggum}, {Taylor}, {Taylor},
  {Turner}, {Unal}, {Vallisneri}, {van Haasteren}, {Vigeland}, {Wahl}, {Wang},
  {Witt}, \& {Young}}]{2023arXiv230616213A}
{Agazie}, G., {Anumarlapudi}, A., {Archibald}, A.~M., {et~al.} 2023, arXiv
  e-prints, arXiv:2306.16213

\bibitem[{{Amaro-Seoane} {et~al.}(2010){Amaro-Seoane}, {Sesana}, {Hoffman},
  {Benacquista}, {Eichhorn}, {Makino}, \& {Spurzem}}]{2010MNRAS.402.2308A}
{Amaro-Seoane}, P., {Sesana}, A., {Hoffman}, L., {et~al.} 2010, \mnras, 402,
  2308

\bibitem[{{Amaro-Seoane} {et~al.}(2022){Amaro-Seoane}, {Andrews}, {Arca Sedda},
  {Askar}, {Balasov}, {Bartos}, {Bavera}, {Bellovary}, {Berry}, {Berti},
  {Bianchi}, {Blecha}, {Blondin}, {Bogdanovi{\'c}}, {Boissier}, {Bonetti},
  {Bonoli}, {Bortolas}, {Breivik}, {Capelo}, {Caramete}, {Catorini}, {Charisi},
  {Chaty}, {Chen}, {Chru{\'s}li{\'n}ska}, {Chua}, {Church}, {Colpi},
  {D'Orazio}, {Danielski}, {Davies}, {Dayal}, {De Rosa}, {Derdzinski},
  {Destounis}, {Dotti}, {Du{\c{t}}an}, {Dvorkin}, {Fabj}, {Foglizzo}, {Ford},
  {Fouvry}, {Fragkos}, {Fryer}, {Gaspari}, {Gerosa}, {Graziani}, {Groot},
  {Habouzit}, {Haggard}, {Haiman}, {Han}, {Istrate}, {Johansson}, {Khan},
  {Kimpson}, {Kokkotas}, {Kong}, {Korol}, {Kremer}, {Kupfer}, {Lamberts},
  {Larson}, {Lau}, {Liu}, {Lloyd-Ronning}, {Lodato}, {Lupi}, {Ma}, {Maccarone},
  {Mandel}, {Mangiagli}, {Mapelli}, {Mathis}, {Mayer}, {McGee}, {McKernan},
  {Miller}, {Mota}, {Mumpower}, {Nasim}, {Nelemans}, {Noble}, {Pacucci},
  {Panessa}, {Paschalidis}, {Pfister}, {Porquet}, {Quenby}, {R{\"o}pke},
  {Regan}, {Rosswog}, {Ruiter}, {Ruiz}, {Runnoe}, {Schneider}, {Schnittman},
  {Secunda}, {Sesana}, {Seto}, {Shao}, {Shapiro}, {Sopuerta}, {Stone},
  {Suvorov}, {Tamanini}, {Tamfal}, {Tauris}, {Temmink}, {Tomsick}, {Toonen},
  {Torres-Orjuela}, {Toscani}, {Tsokaros}, {Unal}, {V{\'a}zquez-Aceves},
  {Valiante}, {van Putten}, {van Roestel}, {Vignali}, {Volonteri}, {Wu},
  {Younsi}, {Yu}, {Zane}, {Zwick}, {Antonini}, {Baibhav}, {Barausse}, {Bonilla
  Rivera}, {Branchesi}, {Branduardi-Raymont}, {Burdge}, {Chakraborty},
  {Cuadra}, {Dage}, {Davis}, {de Mink}, {Decarli}, {Doneva}, {Escoffier},
  {Gandhi}, {Haardt}, {Lousto}, {Nissanke}, {Nordhaus}, {O'Shaughnessy},
  {Portegies Zwart}, {Pound}, {Schussler}, {Sergijenko}, {Spallicci},
  {Vernieri}, \& {Vigna-G{\'o}mez}}]{2022arXiv220306016A}
{Amaro-Seoane}, P., {Andrews}, J., {Arca Sedda}, M., {et~al.} 2022, arXiv
  e-prints, arXiv:2203.06016

\bibitem[{{Artymowicz} \& {Lubow}(1996)}]{1996ApJ...467L..77A}
{Artymowicz}, P., \& {Lubow}, S.~H. 1996, \apjl, 467, L77

\bibitem[{{Avara} {et~al.}(2023){Avara}, {Krolik}, {Campanelli}, {Noble},
  {Bowen}, \& {Ryu}}]{2023arXiv230518538A}
{Avara}, M.~J., {Krolik}, J.~H., {Campanelli}, M., {et~al.} 2023, arXiv
  e-prints, arXiv:2305.18538

\bibitem[{{Bate} \& {Bonnell}(1997)}]{1997MNRAS.285...33B}
{Bate}, M.~R., \& {Bonnell}, I.~A. 1997, \mnras, 285, 33

\bibitem[{{Begelman} {et~al.}(1980){Begelman}, {Blandford}, \&
  {Rees}}]{1980Natur.287..307B}
{Begelman}, M.~C., {Blandford}, R.~D., \& {Rees}, M.~J. 1980, \nat, 287, 307

\bibitem[{{Bellovary} {et~al.}(2019){Bellovary}, {Cleary}, {Munshi}, {Tremmel},
  {Christensen}, {Brooks}, \& {Quinn}}]{2019MNRAS.482.2913B}
{Bellovary}, J.~M., {Cleary}, C.~E., {Munshi}, F., {et~al.} 2019, \mnras, 482,
  2913

\bibitem[{{Bondi}(1952)}]{1952MNRAS.112..195B}
{Bondi}, H. 1952, \mnras, 112, 195

\bibitem[{{Boss}(1986)}]{1986ApJS...62..519B}
{Boss}, A.~P. 1986, \apjs, 62, 519

\bibitem[{{Bowen} {et~al.}(2019){Bowen}, {Mewes}, {Noble}, {Avara},
  {Campanelli}, \& {Krolik}}]{2019ApJ...879...76B}
{Bowen}, D.~B., {Mewes}, V., {Noble}, S.~C., {et~al.} 2019, \apj, 879, 76

\bibitem[{{Bowler} \& {Nielsen}(2018)}]{2018haex.bookE.155B}
{Bowler}, B.~P., \& {Nielsen}, E.~L. 2018, in Handbook of Exoplanets, ed. H.~J.
  {Deeg} \& J.~A. {Belmonte}, 155

\bibitem[{{Bryan} {et~al.}(2018){Bryan}, {Benneke}, {Knutson}, {Batygin}, \&
  {Bowler}}]{2018NatAs...2..138B}
{Bryan}, M.~L., {Benneke}, B., {Knutson}, H.~A., {Batygin}, K., \& {Bowler},
  B.~P. 2018, Nature Astronomy, 2, 138

\bibitem[{{Charisi} {et~al.}(2016){Charisi}, {Bartos}, {Haiman},
  {Price-Whelan}, {Graham}, {Bellm}, {Laher}, \&
  {M{\'a}rka}}]{2016MNRAS.463.2145C}
{Charisi}, M., {Bartos}, I., {Haiman}, Z., {et~al.} 2016, \mnras, 463, 2145

\bibitem[{{Chauvin} {et~al.}(2004){Chauvin}, {Lagrange}, {Dumas}, {Zuckerman},
  {Mouillet}, {Song}, {Beuzit}, \& {Lowrance}}]{2004A&A...425L..29C}
{Chauvin}, G., {Lagrange}, A.~M., {Dumas}, C., {et~al.} 2004, \aap, 425, L29

\bibitem[{{Chen} {et~al.}(2020){Chen}, {Liu}, {Liao}, {Holgado}, {Guo},
  {Gruendl}, {Morganson}, {Shen}, {Zhang}, {Abbott}, {Aguena}, {Allam},
  {Avila}, {Bertin}, {Bhargava}, {Brooks}, {Burke}, {Carnero Rosell},
  {Carollo}, {Carrasco Kind}, {Carretero}, {Costanzi}, {da Costa}, {Davis}, {De
  Vicente}, {Desai}, {Diehl}, {Doel}, {Everett}, {Flaugher}, {Friedel},
  {Frieman}, {Garc{\'\i}a-Bellido}, {Gaztanaga}, {Glazebrook}, {Gruen},
  {Gutierrez}, {Hinton}, {Hollowood}, {James}, {Kim}, {Kuehn}, {Kuropatkin},
  {Lewis}, {Lidman}, {Lima}, {Maia}, {March}, {Marshall}, {Menanteau},
  {Miquel}, {Palmese}, {Paz-Chinch{\'o}n}, {Plazas}, {Sanchez}, {Schubnell},
  {Serrano}, {Sevilla-Noarbe}, {Smith}, {Suchyta}, {Swanson}, {Tarle},
  {Tucker}, {Norbert Varga}, \& {Walker}}]{2020MNRAS.499.2245C}
{Chen}, Y.-C., {Liu}, X., {Liao}, W.-T., {et~al.} 2020, \mnras, 499, 2245

\bibitem[{{Choksi} {et~al.}(2023){Choksi}, {Chiang}, {Fung}, \&
  {Zhu}}]{2023arXiv230501684C}
{Choksi}, N., {Chiang}, E., {Fung}, J., \& {Zhu}, Z. 2023, arXiv e-prints,
  arXiv:2305.01684

\bibitem[{{Cimerman} \& {Rafikov}(2023)}]{2023arXiv230801967C}
{Cimerman}, N.~P., \& {Rafikov}, R.~R. 2023, arXiv e-prints, arXiv:2308.01967

\bibitem[{{Combi} {et~al.}(2022){Combi}, {Lopez Armengol}, {Campanelli},
  {Noble}, {Avara}, {Krolik}, \& {Bowen}}]{2022ApJ...928..187C}
{Combi}, L., {Lopez Armengol}, F.~G., {Campanelli}, M., {et~al.} 2022, \apj,
  928, 187

\bibitem[{{Danby}(1988)}]{1988fcm..book.....D}
{Danby}, J.~M.~A. 1988, {Fundamentals of celestial mechanics}

\bibitem[{{Dempsey} {et~al.}(2020){Dempsey}, {Mu{\~n}oz}, \&
  {Lithwick}}]{2020ApJ...892L..29D}
{Dempsey}, A.~M., {Mu{\~n}oz}, D., \& {Lithwick}, Y. 2020, \apjl, 892, L29

\bibitem[{{Dempsey} {et~al.}(2021){Dempsey}, {Mu{\~n}oz}, \&
  {Lithwick}}]{2021ApJ...918L..36D}
{Dempsey}, A.~M., {Mu{\~n}oz}, D.~J., \& {Lithwick}, Y. 2021, \apjl, 918, L36

\bibitem[{{Dittmann} \& {Miller}(2020)}]{2020MNRAS.493.3732D}
{Dittmann}, A.~J., \& {Miller}, M.~C. 2020, \mnras, 493, 3732

\bibitem[{{Dittmann} \& {Ryan}(2021)}]{2021ApJ...921...71D}
{Dittmann}, A.~J., \& {Ryan}, G. 2021, \apj, 921, 71

\bibitem[{{Dittmann} \& {Ryan}(2022)}]{2022MNRAS.513.6158D}
---. 2022, \mnras, 513, 6158

\bibitem[{{Dittmann} {et~al.}(2023){Dittmann}, {Ryan}, \&
  {Miller}}]{2023ApJ...949L..30D}
{Dittmann}, A.~J., {Ryan}, G., \& {Miller}, M.~C. 2023, \apjl, 949, L30

\bibitem[{{D'Orazio} {et~al.}(2016){D'Orazio}, {Haiman}, {Duffell},
  {MacFadyen}, \& {Farris}}]{2016MNRAS.459.2379D}
{D'Orazio}, D.~J., {Haiman}, Z., {Duffell}, P., {MacFadyen}, A., \& {Farris},
  B. 2016, \mnras, 459, 2379

\bibitem[{{D'Orazio} {et~al.}(2013){D'Orazio}, {Haiman}, \&
  {MacFadyen}}]{2013MNRAS.436.2997D}
{D'Orazio}, D.~J., {Haiman}, Z., \& {MacFadyen}, A. 2013, \mnras, 436, 2997

\bibitem[{{Duffell}(2016)}]{2016ApJS..226....2D}
{Duffell}, P.~C. 2016, \apjs, 226, 2

\bibitem[{{Duffell} {et~al.}(2020){Duffell}, {D'Orazio}, {Derdzinski},
  {Haiman}, {MacFadyen}, {Rosen}, \& {Zrake}}]{2020ApJ...901...25D}
{Duffell}, P.~C., {D'Orazio}, D., {Derdzinski}, A., {et~al.} 2020, \apj, 901,
  25

\bibitem[{{El-Badry} {et~al.}(2019){El-Badry}, {Rix}, {Tian}, {Duch{\^e}ne}, \&
  {Moe}}]{2019MNRAS.489.5822E}
{El-Badry}, K., {Rix}, H.-W., {Tian}, H., {Duch{\^e}ne}, G., \& {Moe}, M. 2019,
  \mnras, 489, 5822

\bibitem[{{Farris} {et~al.}(2014){Farris}, {Duffell}, {MacFadyen}, \&
  {Haiman}}]{2014ApJ...783..134F}
{Farris}, B.~D., {Duffell}, P., {MacFadyen}, A.~I., \& {Haiman}, Z. 2014, \apj,
  783, 134

\bibitem[{{Felker} \& {Stone}(2018)}]{2018JCoPh.375.1365F}
{Felker}, K.~G., \& {Stone}, J.~M. 2018, Journal of Computational Physics, 375,
  1365

\bibitem[{{Franchini} {et~al.}(2021){Franchini}, {Sesana}, \&
  {Dotti}}]{2021MNRAS.507.1458F}
{Franchini}, A., {Sesana}, A., \& {Dotti}, M. 2021, \mnras, 507, 1458

\bibitem[{{Gammie}(2001)}]{2001ApJ...553..174G}
{Gammie}, C.~F. 2001, \apj, 553, 174

\bibitem[{{Gilbaum} \& {Stone}(2022)}]{2022ApJ...928..191G}
{Gilbaum}, S., \& {Stone}, N.~C. 2022, \apj, 928, 191

\bibitem[{{Goldreich} \& {Tremaine}(1979)}]{1979ApJ...233..857G}
{Goldreich}, P., \& {Tremaine}, S. 1979, \apj, 233, 857

\bibitem[{{Graham} {et~al.}(2015){Graham}, {Djorgovski}, {Stern}, {Drake},
  {Mahabal}, {Donalek}, {Glikman}, {Larson}, \&
  {Christensen}}]{2015MNRAS.453.1562G}
{Graham}, M.~J., {Djorgovski}, S.~G., {Stern}, D., {et~al.} 2015, \mnras, 453,
  1562

\bibitem[{{Haehnelt} \& {Kauffmann}(2000)}]{2000MNRAS.318L..35H}
{Haehnelt}, M.~G., \& {Kauffmann}, G. 2000, \mnras, 318, L35

\bibitem[{{Haiman} {et~al.}(2009){Haiman}, {Kocsis}, \&
  {Menou}}]{2009ApJ...700.1952H}
{Haiman}, Z., {Kocsis}, B., \& {Menou}, K. 2009, \apj, 700, 1952

\bibitem[{{Hanawa} {et~al.}(2010){Hanawa}, {Ochi}, \&
  {Ando}}]{2010ApJ...708..485H}
{Hanawa}, T., {Ochi}, Y., \& {Ando}, K. 2010, \apj, 708, 485

\bibitem[{{Hayasaki} {et~al.}(2007){Hayasaki}, {Mineshige}, \&
  {Sudou}}]{2007PASJ...59..427H}
{Hayasaki}, K., {Mineshige}, S., \& {Sudou}, H. 2007, \pasj, 59, 427

\bibitem[{{Hunter}(2007)}]{4160265}
{Hunter}, J.~D. 2007, Computing in Science Engineering, 9, 90

\bibitem[{{Hwang} {et~al.}(2022){Hwang}, {El-Badry}, {Rix}, {Hamilton}, {Ting},
  \& {Zakamska}}]{2022ApJ...933L..32H}
{Hwang}, H.-C., {El-Badry}, K., {Rix}, H.-W., {et~al.} 2022, \apjl, 933, L32

\bibitem[{{Kanagawa} {et~al.}(2018){Kanagawa}, {Tanaka}, \&
  {Szuszkiewicz}}]{2018ApJ...861..140K}
{Kanagawa}, K.~D., {Tanaka}, H., \& {Szuszkiewicz}, E. 2018, \apj, 861, 140

\bibitem[{{Khan} {et~al.}(2013){Khan}, {Holley-Bockelmann}, {Berczik}, \&
  {Just}}]{2013ApJ...773..100K}
{Khan}, F.~M., {Holley-Bockelmann}, K., {Berczik}, P., \& {Just}, A. 2013,
  \apj, 773, 100

\bibitem[{{Kley} \& {Dirksen}(2006)}]{2006A&A...447..369K}
{Kley}, W., \& {Dirksen}, G. 2006, \aap, 447, 369

\bibitem[{{Krauth} {et~al.}(2023){Krauth}, {Davelaar}, {Haiman},
  {Westernacher-Schneider}, {Zrake}, \& {MacFadyen}}]{2023arXiv230402575M}
{Krauth}, L.~M., {Davelaar}, J., {Haiman}, Z., {et~al.} 2023, arXiv e-prints,
  arXiv:2304.02575

\bibitem[{{Kurganov} \& {Tadmor}(2000)}]{2000JCoPh.160..241K}
{Kurganov}, A., \& {Tadmor}, E. 2000, Journal of Computational Physics, 160,
  241

\bibitem[{{Li} {et~al.}(2021){Li}, {Chen}, {Lin}, \&
  {Zhang}}]{2021ApJ...906...52L}
{Li}, Y.-P., {Chen}, Y.-X., {Lin}, D. N.~C., \& {Zhang}, X. 2021, \apj, 906, 52

\bibitem[{{Li} {et~al.}(2022){Li}, {Dempsey}, {Li}, {Li}, \&
  {Li}}]{2022ApJ...928L..19L}
{Li}, Y.-P., {Dempsey}, A.~M., {Li}, H., {Li}, S., \& {Li}, J. 2022, \apjl,
  928, L19

\bibitem[{{Lin} \& {Papaloizou}(1986)}]{1986ApJ...309..846L}
{Lin}, D.~N.~C., \& {Papaloizou}, J. 1986, \apj, 309, 846

\bibitem[{{Liu} {et~al.}(2018){Liu}, {Gezari}, \&
  {Miller}}]{2018ApJ...859L..12L}
{Liu}, T., {Gezari}, S., \& {Miller}, M.~C. 2018, \apjl, 859, L12

\bibitem[{{Liu} {et~al.}(2016){Liu}, {Gezari}, {Burgett}, {Chambers}, {Draper},
  {Hodapp}, {Huber}, {Kudritzki}, {Magnier}, {Metcalfe}, {Tonry}, {Wainscoat},
  \& {Waters}}]{2016ApJ...833....6L}
{Liu}, T., {Gezari}, S., {Burgett}, W., {et~al.} 2016, \apj, 833, 6

\bibitem[{{MacFadyen} \& {Milosavljevi{\'c}}(2008)}]{2008ApJ...672...83M}
{MacFadyen}, A.~I., \& {Milosavljevi{\'c}}, M. 2008, \apj, 672, 83

\bibitem[{{Mahesh} {et~al.}(2023){Mahesh}, {McWilliams}, \&
  {Pirog}}]{2023arXiv230501533M}
{Mahesh}, S., {McWilliams}, S.~T., \& {Pirog}, M. 2023, arXiv e-prints,
  arXiv:2305.01533

\bibitem[{{Miranda} {et~al.}(2017){Miranda}, {Mu{\~n}oz}, \&
  {Lai}}]{2017MNRAS.466.1170M}
{Miranda}, R., {Mu{\~n}oz}, D.~J., \& {Lai}, D. 2017, \mnras, 466, 1170

\bibitem[{{Moody} {et~al.}(2019){Moody}, {Shi}, \&
  {Stone}}]{2019ApJ...875...66M}
{Moody}, M. S.~L., {Shi}, J.-M., \& {Stone}, J.~M. 2019, \apj, 875, 66

\bibitem[{{Mu{\~n}oz} {et~al.}(2020){Mu{\~n}oz}, {Lai}, {Kratter}, \&
  {Miranda}}]{2020ApJ...889..114M}
{Mu{\~n}oz}, D.~J., {Lai}, D., {Kratter}, K., \& {Miranda}, R. 2020, \apj, 889,
  114

\bibitem[{{Mu{\~n}oz} {et~al.}(2019){Mu{\~n}oz}, {Miranda}, \&
  {Lai}}]{2019ApJ...871...84M}
{Mu{\~n}oz}, D.~J., {Miranda}, R., \& {Lai}, D. 2019, \apj, 871, 84

\bibitem[{{M{\"u}ller} {et~al.}(2012){M{\"u}ller}, {Kley}, \&
  {Meru}}]{2012A&A...541A.123M}
{M{\"u}ller}, T.~W.~A., {Kley}, W., \& {Meru}, F. 2012, \aap, 541, A123

\bibitem[{{Narayan} \& {Yi}(1995)}]{1995ApJ...452..710N}
{Narayan}, R., \& {Yi}, I. 1995, \apj, 452, 710

\bibitem[{{Nielsen} {et~al.}(2019){Nielsen}, {De Rosa}, {Macintosh}, {Wang},
  {Ruffio}, {Chiang}, {Marley}, {Saumon}, {Savransky}, {Ammons}, {Bailey},
  {Barman}, {Blain}, {Bulger}, {Burrows}, {Chilcote}, {Cotten}, {Czekala},
  {Doyon}, {Duch{\^e}ne}, {Esposito}, {Fabrycky}, {Fitzgerald}, {Follette},
  {Fortney}, {Gerard}, {Goodsell}, {Graham}, {Greenbaum}, {Hibon}, {Hinkley},
  {Hirsch}, {Hom}, {Hung}, {Dawson}, {Ingraham}, {Kalas}, {Konopacky},
  {Larkin}, {Lee}, {Lin}, {Maire}, {Marchis}, {Marois}, {Metchev},
  {Millar-Blanchaer}, {Morzinski}, {Oppenheimer}, {Palmer}, {Patience},
  {Perrin}, {Poyneer}, {Pueyo}, {Rafikov}, {Rajan}, {Rameau}, {Rantakyr{\"o}},
  {Ren}, {Schneider}, {Sivaramakrishnan}, {Song}, {Soummer}, {Tallis},
  {Thomas}, {Ward-Duong}, \& {Wolff}}]{2019AJ....158...13N}
{Nielsen}, E.~L., {De Rosa}, R.~J., {Macintosh}, B., {et~al.} 2019, \aj, 158,
  13

\bibitem[{{Noble} {et~al.}(2021){Noble}, {Krolik}, {Campanelli}, {Zlochower},
  {Mundim}, {Nakano}, \& {Zilh{\~a}o}}]{2021ApJ...922..175N}
{Noble}, S.~C., {Krolik}, J.~H., {Campanelli}, M., {et~al.} 2021, \apj, 922,
  175

\bibitem[{{Noble} {et~al.}(2012){Noble}, {Mundim}, {Nakano}, {Krolik},
  {Campanelli}, {Zlochower}, \& {Yunes}}]{2012ApJ...755...51N}
{Noble}, S.~C., {Mundim}, B.~C., {Nakano}, H., {et~al.} 2012, \apj, 755, 51

\bibitem[{{Ochi} {et~al.}(2005){Ochi}, {Sugimoto}, \&
  {Hanawa}}]{2005ApJ...623..922O}
{Ochi}, Y., {Sugimoto}, K., \& {Hanawa}, T. 2005, \apj, 623, 922

\bibitem[{{Paardekooper} \& {Papaloizou}(2009)}]{2009MNRAS.394.2283P}
{Paardekooper}, S.~J., \& {Papaloizou}, J.~C.~B. 2009, \mnras, 394, 2283

\bibitem[{{Park} {et~al.}(2022){Park}, {Ricotti}, \&
  {Sugimura}}]{2022arXiv221204564P}
{Park}, J., {Ricotti}, M., \& {Sugimura}, K. 2022, arXiv e-prints,
  arXiv:2212.04564

\bibitem[{{Penzlin} {et~al.}(2022){Penzlin}, {Kley}, {Audiffren}, \&
  {Sch{\"a}fer}}]{2022A&A...660A.101P}
{Penzlin}, A. B.~T., {Kley}, W., {Audiffren}, H., \& {Sch{\"a}fer}, C.~M. 2022,
  \aap, 660, A101

\bibitem[{{Phinney}(2001)}]{2001astro.ph..8028P}
{Phinney}, E.~S. 2001, arXiv e-prints, astro

\bibitem[{{Piso} \& {Youdin}(2014)}]{2014ApJ...786...21P}
{Piso}, A.-M.~A., \& {Youdin}, A.~N. 2014, \apj, 786, 21

\bibitem[{{Rafikov}(2015)}]{2015ApJ...804...62R}
{Rafikov}, R.~R. 2015, \apj, 804, 62

\bibitem[{{Ragusa} {et~al.}(2020){Ragusa}, {Alexander}, {Calcino}, {Hirsh}, \&
  {Price}}]{2020MNRAS.499.3362R}
{Ragusa}, E., {Alexander}, R., {Calcino}, J., {Hirsh}, K., \& {Price}, D.~J.
  2020, \mnras, 499, 3362

\bibitem[{{Rajagopal} \& {Romani}(1995)}]{1995ApJ...446..543R}
{Rajagopal}, M., \& {Romani}, R.~W. 1995, \apj, 446, 543

\bibitem[{{Reed} {et~al.}(2007){Reed}, {Bower}, {Frenk}, {Jenkins}, \&
  {Theuns}}]{2007MNRAS.374....2R}
{Reed}, D.~S., {Bower}, R., {Frenk}, C.~S., {Jenkins}, A., \& {Theuns}, T.
  2007, \mnras, 374, 2

\bibitem[{{Roe}(1981)}]{1981JCoPh..43..357R}
{Roe}, P.~L. 1981, Journal of Computational Physics, 43, 357

\bibitem[{{Rosenthal} {et~al.}(2020){Rosenthal}, {Chiang}, {Ginzburg}, \&
  {Murray-Clay}}]{2020MNRAS.498.2054R}
{Rosenthal}, M.~M., {Chiang}, E.~I., {Ginzburg}, S., \& {Murray-Clay}, R.~A.
  2020, \mnras, 498, 2054

\bibitem[{{Shakura} \& {Sunyaev}(1973)}]{1973A&A....24..337S}
{Shakura}, N.~I., \& {Sunyaev}, R.~A. 1973, \aap, 500, 33

\bibitem[{{Shen} {et~al.}(2008){Shen}, {Greene}, {Strauss}, {Richards}, \&
  {Schneider}}]{2008ApJ...680..169S}
{Shen}, Y., {Greene}, J.~E., {Strauss}, M.~A., {Richards}, G.~T., \&
  {Schneider}, D.~P. 2008, \apj, 680, 169

\bibitem[{{Shi} \& {Krolik}(2015)}]{2015ApJ...807..131S}
{Shi}, J.-M., \& {Krolik}, J.~H. 2015, \apj, 807, 131

\bibitem[{{Shi} {et~al.}(2012){Shi}, {Krolik}, {Lubow}, \&
  {Hawley}}]{2012ApJ...749..118S}
{Shi}, J.-M., {Krolik}, J.~H., {Lubow}, S.~H., \& {Hawley}, J.~F. 2012, \apj,
  749, 118

\bibitem[{{Sirko} \& {Goodman}(2003)}]{2003MNRAS.341..501S}
{Sirko}, E., \& {Goodman}, J. 2003, \mnras, 341, 501

\bibitem[{{Siwek} {et~al.}(2023){Siwek}, {Weinberger}, {Mu{\~n}oz}, \&
  {Hernquist}}]{2022arXiv220302514S}
{Siwek}, M., {Weinberger}, R., {Mu{\~n}oz}, D.~J., \& {Hernquist}, L. 2023,
  \mnras, 518, 5059

\bibitem[{{Siwek} {et~al.}(2020){Siwek}, {Kelley}, \&
  {Hernquist}}]{2020MNRAS.498..537S}
{Siwek}, M.~S., {Kelley}, L.~Z., \& {Hernquist}, L. 2020, \mnras, 498, 537

\bibitem[{{S{\k{a}}dowski} {et~al.}(2011){S{\k{a}}dowski}, {Abramowicz},
  {Bursa}, {Klu{\'z}niak}, {Lasota}, \&
  {R{\'o}{\.z}a{\'n}ska}}]{2011A&A...527A..17S}
{S{\k{a}}dowski}, A., {Abramowicz}, M., {Bursa}, M., {et~al.} 2011, \aap, 527,
  A17

\bibitem[{{Stone} \& {Gardiner}(2009)}]{2009NewA...14..139S}
{Stone}, J.~M., \& {Gardiner}, T. 2009, \na, 14, 139

\bibitem[{{Stone} {et~al.}(2020){Stone}, {Tomida}, {White}, \&
  {Felker}}]{2020ApJS..249....4S}
{Stone}, J.~M., {Tomida}, K., {White}, C.~J., \& {Felker}, K.~G. 2020, \apjs,
  249, 4

\bibitem[{{Taffoni} {et~al.}(2003){Taffoni}, {Mayer}, {Colpi}, \&
  {Governato}}]{2003MNRAS.341..434T}
{Taffoni}, G., {Mayer}, L., {Colpi}, M., \& {Governato}, F. 2003, \mnras, 341,
  434

\bibitem[{{Tamfal} {et~al.}(2018){Tamfal}, {Capelo}, {Kazantzidis}, {Mayer},
  {Potter}, {Stadel}, \& {Widrow}}]{2018ApJ...864L..19T}
{Tamfal}, T., {Capelo}, P.~R., {Kazantzidis}, S., {et~al.} 2018, \apjl, 864,
  L19

\bibitem[{{Tanaka} {et~al.}(2002){Tanaka}, {Takeuchi}, \&
  {Ward}}]{2002ApJ...565.1257T}
{Tanaka}, H., {Takeuchi}, T., \& {Ward}, W.~R. 2002, \apj, 565, 1257

\bibitem[{Thyng {et~al.}(2016)Thyng, Greene, Hetland, Zimmerle, \&
  DiMarco}]{cmocean}
Thyng, K.~M., Greene, C.~A., Hetland, R.~D., Zimmerle, H.~M., \& DiMarco, S.~F.
  2016, Oceanography.
\newblock \url{https://doi.org/10.5670/oceanog.2016.66}

\bibitem[{{Tiede} {et~al.}(2020){Tiede}, {Zrake}, {MacFadyen}, \&
  {Haiman}}]{2020ApJ...900...43T}
{Tiede}, C., {Zrake}, J., {MacFadyen}, A., \& {Haiman}, Z. 2020, \apj, 900, 43

\bibitem[{{Tiede} {et~al.}(2022){Tiede}, {Zrake}, {MacFadyen}, \&
  {Haiman}}]{2022ApJ...932...24T}
---. 2022, \apj, 932, 24

\bibitem[{{Turk} {et~al.}(2011){Turk}, {Smith}, {Oishi}, {Skory}, {Skillman},
  {Abel}, \& {Norman}}]{2011ApJS..192....9T}
{Turk}, M.~J., {Smith}, B.~D., {Oishi}, J.~S., {et~al.} 2011, \apjs, 192, 9

\bibitem[{{Turner} \& {Reynolds}(2023)}]{2023arXiv230607199T}
{Turner}, S. G.~D., \& {Reynolds}, C.~S. 2023, arXiv e-prints, arXiv:2306.07199

\bibitem[{{van der Walt} {et~al.}(2011){van der Walt}, {Colbert}, \&
  {Varoquaux}}]{5725236}
{van der Walt}, S., {Colbert}, S.~C., \& {Varoquaux}, G. 2011, Computing in
  Science Engineering, 13, 22

\bibitem[{{Vaughan} {et~al.}(2016){Vaughan}, {Uttley}, {Markowitz},
  {Huppenkothen}, {Middleton}, {Alston}, {Scargle}, \&
  {Farr}}]{2016MNRAS.461.3145V}
{Vaughan}, S., {Uttley}, P., {Markowitz}, A.~G., {et~al.} 2016, \mnras, 461,
  3145

\bibitem[{{Velikhov}(1959)}]{velikhov59}
{Velikhov}, E. 1959, JETP, 36, 1398

\bibitem[{{Wagner} {et~al.}(2019){Wagner}, {Apai}, \&
  {Kratter}}]{2019ApJ...877...46W}
{Wagner}, K., {Apai}, D., \& {Kratter}, K.~M. 2019, \apj, 877, 46

\bibitem[{{Westernacher-Schneider} {et~al.}(2022){Westernacher-Schneider},
  {Zrake}, {MacFadyen}, \& {Haiman}}]{2022PhRvD.106j3010W}
{Westernacher-Schneider}, J.~R., {Zrake}, J., {MacFadyen}, A., \& {Haiman}, Z.
  2022, \prd, 106, 103010

\bibitem[{{Young} {et~al.}(2015){Young}, {Baird}, \&
  {Clarke}}]{2015MNRAS.447.2907Y}
{Young}, M.~D., {Baird}, J.~T., \& {Clarke}, C.~J. 2015, \mnras, 447, 2907

\bibitem[{{Zrake} {et~al.}(2021){Zrake}, {Tiede}, {MacFadyen}, \&
  {Haiman}}]{2020arXiv201009707Z}
{Zrake}, J., {Tiede}, C., {MacFadyen}, A., \& {Haiman}, Z. 2021, \apjl, 909,
  L13

\end{thebibliography}
\end{document}